\documentclass[
reprint, 
twocolumn,
superscriptaddress, amsmath, amssymb, aps, pre]{revtex4-2}
\usepackage{lineno}
\usepackage{graphicx}
\usepackage{dcolumn}
\usepackage{bm}

\begin{document}
\title{Improving atomic force microscopy structure discovery via style-translation}

\author{Jie Huang}
\affiliation{Department of Applied Physics, Aalto University, Helsinki, FI-02150, Finland}

\author{Niko Oinonen}
\affiliation{Department of Applied Physics, Aalto University, Helsinki, FI-02150, Finland}
\affiliation{Nanolayers Research Computing Ltd., 51 New Way Road, London, NW9 6PL, United Kingdom}

\author{Fabio Priante}
\affiliation{Department of Chemistry and Materials Science, Aalto University, Helsinki, FI-02150, Finland}

\author{Filippo Federici Canova}
\affiliation{Department of Applied Physics, Aalto University, Helsinki, FI-02150, Finland}
\affiliation{Nanolayers Research Computing Ltd., London N12 0HL, United Kingdom}

\author{Lauri Kurki}
\affiliation{Department of Applied Physics, Aalto University, Helsinki, FI-02150, Finland}

\author{Chen Xu}
\affiliation{Department of Applied Physics, Aalto University, Helsinki, FI-02150, Finland}

\author{Adam S. Foster}
\email{adam.foster@aalto.fi}
\affiliation{Department of Applied Physics, Aalto University, Helsinki, FI-02150, Finland}
\affiliation{WPI Nano Life Science Institute (WPI-NanoLSI), Kanazawa University, Kakuma-machi, Kanazawa, 920-1192, Japan}

\date{\today}
\begin{abstract}
Atomic force microscopy (AFM) is a key tool for characterising nanoscale structures, with functionalised tips now offering detailed images of the atomic structure. In parallel, AFM simulations using the particle probe model provide a cost-effective approach for rapid AFM image generation. Using state-of-the-art machine learning models and substantial simulated datasets, properties such as molecular structure, electrostatic potential, and molecular graph can be predicted from AFM images. However, transferring model performance from simulated to experimental AFM images poses challenges due to the subtle variations in real experimental data compared to the seemingly flawless simulations. In this study, we explore style translation to augment simulated images and improve the predictive performance of machine learning models in surface property analysis. We reduce the style gap between simulated and experimental AFM images and demonstrate the method's effectiveness in enhancing structure discovery models through local structural property distribution comparisons. This research presents a novel approach to improving the efficiency of machine learning models in the absence of labelled experimental data.
\end{abstract}
\maketitle

\section{Introduction}
Understanding material design and properties at the nanoscale relies heavily on visualising atomic structures. Among many characterisation tools, atomic force microscopy (AFM) stands out for its ability to image surfaces with atomic resolution \cite{Giessibl2003, Giessibl2024}. In particular, frequency-modulated non-contact AFM provides contrast that reflects tip-sample interactions, enabling indirect visualisation of atomic configurations through frequency shift signals \cite{Martin1987, GARCIA2002, Barth2010}. However, interpreting AFM images, especially for complex molecules, requires both experience and intuition \cite{Gross2010, Albrecht2015}. While experts can propose reasonable structure hypotheses, validating these guesses typically involves density functional theory (DFT) calculations and AFM simulations like the probe particle model (PPM) \cite{Hapala2014PRB, Hapala2014PRL, Oinonen2024} to compare simulated AFM images with experiments \cite{RaheCalcite2023, Cai2025}. This simulation-aided process has proven valuable, but it is iterative, computationally demanding, and challenging to scale for large systems.  Consequently, it is appealing to develop machine learning (ML) models that are capable of directly predicting and inferring atomic configurations from AFM images, thereby possibly automating the structure discovery process.

Recent advances in ML have shown promise for automating structure discovery from scanning probe microscopy images. A wide range of methods has been explored. Convolutional neural networks (CNNs) have been used to map AFM images to atomic descriptors such as van der Waals (vdW) spheres \cite{BenjaminA2020}, and to predict atomic positions and orientations in interfacial ionic hydrates \cite{Tang2022}. CNN-based approaches have also been applied to automated molecular recognition in scanning tunnelling microscope (STM) images \cite{Zhu2022, Kurki2024}.  By integrating CNNs with graph neural networks (GNNs), it is possible to predict complete atomic graphs, including 3D coordinates, directly from AFM data \cite{Oinonen2022, Priante2024}.  Other approaches leverage AFM image fingerprints to identify molecular candidates and assign confidence scores to predictions \cite{GonzlezLastre2024}. In parallel, multimodal recurrent neural networks (mRNNs) have been used to infer chemical nomenclature, such as IUPAC names, from AFM images \cite{CarracedoCosme2023}. Generative models have also contributed to the progress. Variational autoencoders (VAEs) have been used to synthesise AFM images to improve molecular classification tasks \cite{CarracedoCosme2021}. Conditional generative adversarial networks (cGANs) have been used to convert AFM images into interpretable ball-and-stick representations \cite{CarracedoCosme2024b}. More broadly,  GANs have been applied to generate realistic microscopy images, such as in scanning transmission electron microscopy (STEM), for training defect detection models \cite{Khan2023}, and to simulate STM images \cite{Zhu2024}. Additionally,  support vector machines (SVMs) have been used for real-time classification and feature recognition during AFM measurements \cite{Huang2018}.

Despite these advances, two key challenges continue to limit the real-world deployment of ML-based structure discovery tools. First, the simulation-to-real domain gap poses a fundamental obstacle. Although PPM-based simulation can produce AFM images that closely resemble experimental observations, the style mismatch between domains still exists: real AFM images often contain noise, artefacts, and subtle distortions that are absent in idealised simulations. This domain gap hinders the generalisation of simulation-trained models to experimental data. Second, the scarcity of ground-truth atomic structures for experimental AFM images makes it extremely difficult to systematically evaluate the structure prediction performance on real datasets, let alone training the model on experimental data.

To tackle these challenges, we propose an approach that integrates style translation \cite{zhu2020} with structure prediction \cite{BenjaminA2020, Oinonen2022, Priante2024} to bridge the gap between simulated and experimental AFM images. By translating simulated images into experimental-like ones, we improve model performance on real experimental data. Additionally, we introduce structure-based evaluation metrics that do not rely on ground-truth atomic structures. Our results show that style translation enhances prediction accuracy and enables evaluation in the absence of direct validation.

In the next section, we outline the challenges and hypotheses in ML-based structure discovery from AFM images. Section \ref{sec:cycleGAN} introduces our style translation framework. Section \ref{sec:structureDiscovery} describes dataset construction and presents prediction comparisons across models. Finally, Section \ref{sec:performanceEva} details the model performance evaluations based on physically meaningful structural properties.
\section{Problem definitions and hypothesis}\label{sec:problem}
\begin{figure}[ht!]
    \centering
    \includegraphics[width=0.5\textwidth]{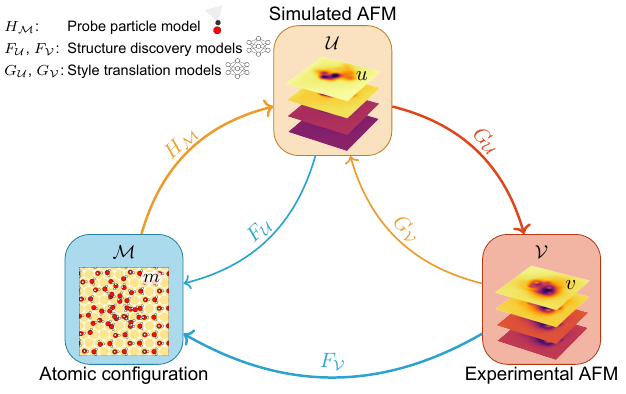}
    \caption{\textbf{Overview of structure discovery from experimental AFM images.} Simulated AFM images $\mathcal{U}$ are generated from atomic configurations $\mathcal{M}$ using the probe particle model (PPM) $H_\mathcal{M}$. A machine learning-based structure discovery model $F_\mathcal{U}$ is trained to recover the atomic configuration from a simulated 3D AFM image, learning the mapping from $\mathcal{U}$ to $\mathcal{M}$. Ideally, the structure discovery model $F_\mathcal{V}$ is trained on experimental style AFM images $\mathcal{V}$ and their corresponding atomic configurations $\mathcal{M}$, enabling structure prediction directly from experimental data. To bridge the style gap between simulated and experimental AFM images, style translators $G_\mathcal{U}$ and $G_\mathcal{V}$ are used to translate domains $\mathcal{U}$ and $\mathcal{V}$, respectively.}
    \label{fig:overview}
\end{figure}

As shown in Fig. \ref{fig:overview}, we let $m$, $u$, $v$ represent an atomic configuration, a simulated 3D AFM image, and an experimental 3D AFM image, respectively. Correspondingly, $\mathcal{M}$, $\mathcal{U}$, and $\mathcal{V}$ denote the domains of atomic configurations, simulated AFM images, and experimental AFM images. The forward problem is defined as: given an atomic configuration, what is the corresponding 3D AFM image? This question can be addressed in two ways. The first is by physically performing the AFM experiment for the given sample, thereby obtaining experimental 3D AFM images, i.e., realising the mapping from $\mathcal{M}$ to $\mathcal{V}$. The second approach involves simulating 3D AFM images from atomic configurations using the PPM, representing the mapping from $\mathcal{M}$ to $\mathcal{U}$.  In contrast, the inverse problem is defined as: given a 3D AFM image, what is the corresponding atomic structure? Assuming that the simulated and experimental AFM images are drawn from the same underlying data distribution, a structure discovery model $F_\mathcal{U}$ is trained using pairs from the domain $\mathcal{U}$ and their corresponding atomic configurations in $\mathcal{M}$. The trained model $F_\mathcal{U}$ is then applied directly to experimental AFM images from domain $\mathcal{V}$. This assumption is primarily driven by two practical limitations: (1) it is costly and time-consuming to collect sufficient experimental AFM samples,  and (2) more critically, it is often unfeasible to obtain the corresponding ground-truth atomic configurations for experimental images, particularly for large or complex samples. As a result, training is conducted using ($\mathcal{U} \rightarrow \mathcal{M}$) rather than directly using experimental data ($\mathcal{V} \rightarrow \mathcal{M}$). 

However, in this study, we do not assume that simulated and experimental AFM images are drawn from the same distribution. While the simulated AFM images resemble experimental ones quite closely, notable discrepancies remain, especially in image noise and artefacts introduced by real-world experimental conditions. We refer to this discrepancy as the style gap between simulated and experimental AFM images. The first hypothesis of this work is that the presence of a style gap degrades the performance of a model $F_\mathcal{U}$ trained solely on simulated data when applied to the experimental AFM images. The second hypothesis is that reducing this style gap in the training data can improve the model's performance on the experimental AFM images, motivating a data-driven style translation approach, which is described in the next section.

\section{Data-driven style translation}\label{sec:cycleGAN}
To reduce the style gap between the simulated and experimental AFM image domains, we employ an image-to-image translation model that takes simulated images as input and generates images in the style of experimental data. This model is inspired by the GAN \cite{Goodfellow_NIPS2014} framework, a class of machine learning models designed to generate new data samples that resemble a target distribution. Unlike conventional GANs, which generate data from random noise vectors, our image-to-image translation model uses source domain images, simulated AFM images in our case, as direct inputs. Specifically, we adopt a cycle-consistent generative adversarial network (CycleGAN) framework \cite{zhu2020, isola2017image}, which extends the GAN architecture by introducing a cycle consistency constraint. CycleGAN has demonstrated effectiveness across a wide range of applications, including statistical tasks such as density estimation and manifold fitting \cite{Liu2021, YauPNAS2024}, medical image synthesis \cite{Sandfort2019, Wang2023},  and high-fidelity image generation in STEM and STM images \cite{Khan2023, Zhu2024}.

\begin{figure}[ht!]
    \centering
    \includegraphics[width=0.5\textwidth]{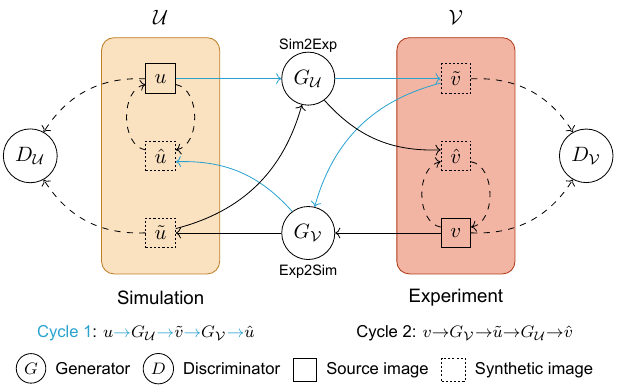}
    \caption{\textbf{Schematic representation of the framework of CycleGAN.} The shaded rectangles indicate the domains of simulated and experimental AFM images, while the circular components represent the individual sub-networks. The solid arrows are the transfer paths within the generators, while the dashed arrows denote the paths for computing the loss function.}
    \label{fig:cyclegan}
\end{figure}

Figure \ref{fig:cyclegan} illustrates the CycleGAN architecture in this study \cite{zhu2020, YauPNAS2024}. We define two domains: $\mathcal{U}$ and $\mathcal{V}$, representing the distributions of simulated and experimental AFM images, respectively. Two datasets are constructed: $\mathcal{U}_M = \{u_i\}_{i=1}^M$ and $\mathcal{V}_N = \{v_i\}_{i=1}^N$, where $u$ and $v$ are 2D AFM image slices, and $M=729$, $N=728$ denote the number of image samples in each dataset. Each 2D AFM slice represents the frequency shift of the cantilever over a horizontal plane at a fixed height. We intentionally set $M \ne N$ to emphasise that the datasets are unpaired, i.e., there is no one-to-one correspondence between images in the two domains. In order to learn the bidirectional mapping between domains, two image-to-image generators are defined: $G_\mathcal{U}: \mathcal{U}\rightarrow\mathcal{V}$ and $G_\mathcal{V}: \mathcal{V}\rightarrow\mathcal{U}$. Given real input images $u$ and $v$, the generators produce the synthetic output points $\tilde{v} = G_\mathcal{U}(u)$ and $\tilde{u} = G_\mathcal{V}(v)$, respectively. To enforce cycle consistency, the outputs are further transformed back to $\hat{u} = G_\mathcal{V}(\tilde{v})$ and $\hat{v} = G_\mathcal{U}(\tilde{u})$, with the aim that $\hat{u} \approx u$ and $\hat{v} \approx v$. Hence, there are two cycle loops as shown in Fig. \ref{fig:cyclegan}.  To distinguish real from generated images, two discriminators are employed: $D_\mathcal{V}$ and $D_\mathcal{U}$. Each discriminator receives an image as input and outputs a number indicating whether the image is real or synthetic. The adversarial training process encourages the generators to produce images indistinguishable from the real domain samples, as judged by the corresponding discriminators.

\noindent\textbf{Adversarial loss.} 
The goal of each generator is to produce synthetic images that resemble real images in the target domain. The adversarial loss function of $G_\mathcal{U}$ and its discriminator $D_\mathcal{V}$ is defined as:
\begin{equation}
\begin{split}
\mathcal{L}_{\mathcal{U}\rightarrow\mathcal{V}}(G_\mathcal{U}, D_\mathcal{V}) & = \frac{1}{n}\sum_{i=1}^{n}\log D_\mathcal{V}(v_i)
\\&\quad 
+ \frac{1}{m}\sum_{i=1}^{m}\log(1 - D_\mathcal{V}(\tilde{v}_i))\\
& = \frac{1}{n}\sum_{i=1}^{n}\log D_\mathcal{V}(v_i)
\\&\quad 
+ \frac{1}{m}\sum_{i=1}^{m}\log(1 - D_\mathcal{V}(G_\mathcal{U}(u_i))).
\end{split}
\end{equation}
Here, $m$, $n$ are the batched sizes of samples from domains $\mathcal{U}$ and $\mathcal{V}$; $\tilde{v}_i = G_\mathcal{U}(u_i)$ is the synthetic output from simulation-to-experiment generator.  The generator $G_\mathcal{U}$ is optimised to minimise the loss, while the discriminator $D_\mathcal{V}$ is trained to maximise it. Similarly, the adversarial loss for $G_\mathcal{V}$ and $D_\mathcal{U}$ is denoted as $\mathcal{L}_{\mathcal{V}\rightarrow\mathcal{U}}(G_\mathcal{V}, D_\mathcal{U})$. 

\noindent\textbf{Cycle consistency loss.} 
To further reduce the space of possible mapping and ensure that translation from one domain and back returns the original image, i.e., $u \to \tilde{v} \to \hat{u} \approx u$ and $v \to \tilde{u} \to \hat{v} \approx v$, CycleGAN introduces a cycle consistency loss:

\begin{equation}
\begin{split}
\mathcal{L}_\text{c} (G_\mathcal{U}, G_\mathcal{V}) & = \frac{1}{m}\sum_{i=0}^{m}\lVert \hat{u}_i - u_i \rVert_1
\\ &\quad 
+ \frac{1}{n}\sum_{i=0}^{n}\lVert \hat{v}_i - v_i \rVert_1, 
\end{split}
\end{equation}
where $\lVert \cdot \rVert_1$ is the $L^1$ norm.

\noindent\textbf{Identity loss.}
To further constrain the generators and prevent unnecessary transformation of images that already resemble the target domain, we introduce an identity loss:

\begin{equation}
\begin{split}
\mathcal{L}_{\text {i}}(G_\mathcal{U}, G_\mathcal{V}) & =
\frac{1}{n}\sum_{i=0}^{n}\lVert G_{\mathcal{U}}(v_i)-v_i\rVert_{1}
\\&\quad 
+ \frac{1}{m}\sum_{i=0}^m\lVert G_\mathcal{V}(u_i)-u_i\rVert_{1}.
\end{split}
\end{equation}
This loss encourages each generator to behave as an identity mapping when provided with inputs from the target domain. In other words, it requires the generator to recognise whether the input image already matches the target style; if so, no translation is needed; otherwise, a style translation is applied. This ensures that translations are applied only when needed, helping avoid over-modification.

\noindent\textbf{Full objective.}
The total loss function used to train CycleGAN is: 

\begin{equation}
\begin{split}
\mathcal{L}\left(G_\mathcal{U}, G_\mathcal{V}, D_\mathcal{V}, D_\mathcal{U}\right) & = \mathcal{L}_{\mathcal{U}\rightarrow\mathcal{V}}(G_\mathcal{U}, D_\mathcal{V}) 
+ \mathcal{L}_{\mathcal{V}\rightarrow\mathcal{U}}(G_\mathcal{V}, D_\mathcal{U}) 
\\&\quad 
+ \lambda_{\text{c}} \mathcal{L}_{\text{c}}(G_\mathcal{U}, G_\mathcal{V}) 
+ \lambda_{\text{i}} \mathcal{L}_{\text{i}}(G_\mathcal{U}, G_\mathcal{V}),
\end{split}
\end{equation}
where $\lambda_{\text{c}}$ and $\lambda_{\text{i}}$ are the weights controlling the contributions of cycle consistency and identity loss, respectively. The optimisation objective is then:
\begin{equation}
G_\mathcal{U}^*, G_\mathcal{V}^* = \arg \min_{G_\mathcal{U}, G_\mathcal{V}} \max_{D_\mathcal{V}, D_\mathcal{U}} \mathcal{L}\left(G_\mathcal{U}, G_\mathcal{V}, D_\mathcal{V}, D_\mathcal{U}\right).
\end{equation}

One of the key advantages of the CycleGAN framework is that it does not require paired images. In the context of AFM, a paired image would consist of simulated and experimental AFM images derived from exactly the same underlying atomic configuration - obtaining sufficient examples in practice is unfeasible. CycleGAN overcomes this limitation by learning from unpaired datasets: it only requires a sufficient number of samples from each domain, without any explicit correspondence between them. After training the CycleGAN framework, we obtain two domain translation models: the forward generator $G_\mathcal{U}$, which maps simulated AFM images from domain $\mathcal{U}$ to the experimental style AFM images in domain $\mathcal{V}$, and the reverse generator $G_\mathcal{V}$, which performs the opposite mapping. 

\begin{figure*}[ht!]
    \centering
    \includegraphics[width=\textwidth]{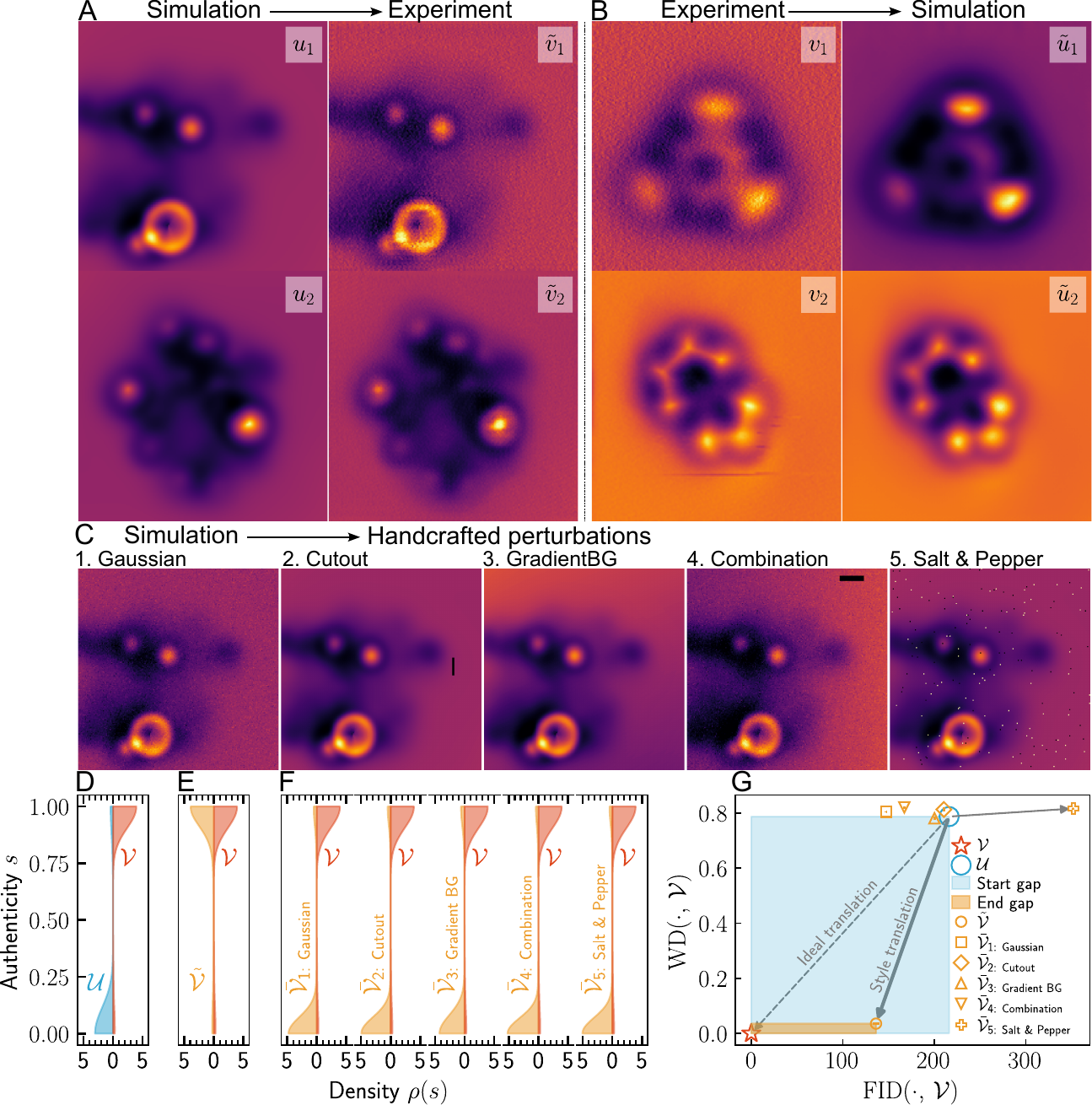}
    \caption{\textbf{Translating AFM image styles and comparing domain shifts using authenticity and distribution metrics.} (\textbf{A}) Simulated AFM images translated into experimental style using generator $G_\mathcal{U}$. (\textbf{B}) Experimental AFM images translated into simulation style using generator $G_\mathcal{V}$. (\textbf{C}) Simulated images with handcrafted perturbations. (\textbf{D})-(\textbf{F}) Authenticity scores $s$ from a machine expert for simulated $\mathcal{U}$, experimental $\mathcal{V}$, experiment style, and perturbed simulated images. (\textbf{G}) Style gap comparisons using Wasserstein distance and Fréchet Inception Distance (FID) relative to the experimental domain $\mathcal{V}$.}
    \label{fig:styleTransResult}
\end{figure*}

\noindent\textbf{Style translation results and style gap evaluations.}
As shown in Fig. \ref{fig:styleTransResult}A, two example 2D simulated AFM images $u_1$ and $u_2$ are transformed by the forward generator $G_\mathcal{U}$ into $\tilde{v}_1, \tilde{v}_2$, respectively, which resemble experimental AFM images. The forward generator learns to add stylistic features like noise into the simulated images to make them visually consistent with experimental images. On the contrary, the reverse generator $G_\mathcal{V}$ produces outputs visually closer to simulated AFM images by removing noise and experimental artefacts, as shown in Fig. \ref{fig:styleTransResult}B, suggesting potential applications in denoising or artefact reduction in experimental images. However, in this study, we focus on the forward translation  ($\mathcal{U} \rightarrow \mathcal{V}$) to reduce the style gap and generate structure discovery training data that better matches the characteristics of experimental AFM images. For comparisons, Fig. \ref{fig:styleTransResult}C  shows several representative handcrafted perturbations for $u_1$, which include  (1) adding Gaussian noises, (2) cutting out random small areas, (3) adding random gradient background, (4) the combinations of the previous three perturbation methods, as well as (5) adding salt and pepper noises. These perturbations, except for (5), are commonly used in previous studies \cite{Kurki2024, Priante2024} as data augmentations.

To evaluate the quality of the style translation, we design a quantitative and data-driven evaluation workflow: 

\noindent\textbf{1. Machine expert training.} We first train a binary classifier, referred to as a machine expert, that takes a 2D AFM image as the input and outputs an authenticity score $s \in [0, 1]$. This classifier is trained on the datasets using two kinds of labelled data: one consisting of simulated AFM images labelled with $s=0$ and another consisting of experimental AFM images labelled with $s=1$. As shown in Fig. \ref{fig:styleTransResult}D, the resulting authenticity score distributions, denoted $\rho_\mathcal{U}(s)$ for simulated data and $\rho_\mathcal{V}(s)$ for experimental data, are well-separated. This indicates that the classifier is effective in distinguishing between the two domains. 

\noindent\textbf{2. Authenticity distribution shift after style translation.} We apply the forward generator $G_\mathcal{U}$ to the set $\mathcal{U}$, producing the style-translated image set $\tilde{\mathcal{V}} = G_\mathcal{U}(\mathcal{U})$. This set is then evaluated using the machine expert to obtain the authenticity distribution $\rho_{\tilde{\mathcal{V}}}(s)$. As shown in Fig. \ref{fig:styleTransResult}E, the distribution $\rho_{\tilde{\mathcal{V}}}(s)$ closely aligns with $\rho_\mathcal{V}(s)$, demonstrating that the forward style generator $G_\mathcal{U}$ successfully bridges the style gap between simulation and experiment. For comparison, Fig. \ref{fig:styleTransResult}F shows the authenticity distributions after applying handcrafted perturbations. These methods fail to produce observable shifts in authenticity, suggesting that these perturbations are less effective in making the simulated images resemble experimental ones, at least from the perspective of the machine expert.


\noindent\textbf{3. Quantitative evaluation with distribution metrics.}
To quantitatively assess the style similarity, we compute the Wasserstein Distance (also known as Earth Mover's Distance) \cite{weng2017gan, arjovsky2017wassersteingan, Herrmann2017EMD} $\mathrm{WD}(\cdot \| \mathcal{V})$ between the authenticity distributions of the translated/perturbed sets and the experimental distributions $\mathcal{V}$. Additionally, we measure the corresponding Fréchet Inception Distance (FID) \cite{HeuselFID2018} $\mathrm{FID}(\cdot \| \mathcal{V})$ between the distributions of different image sets and the distribution of the authentic experimental image set $\mathcal{V}$. Unlike the Wasserstein distance $\mathrm{WD}(\cdot | \mathcal{V})$, which uses features from a lightweight, AFM-specific model (the machine expert), FID is based on features from a large Inception network \cite{szegedy2015Inception} trained on general images from ImageNet \cite{ILSVRC15}. Although FID is not specifically designed for AFM images, it can still capture statistical differences between AFM domains before and after style translation. Both methods have their respective strengths and limitations; hence,  we use both metrics with the motivation to provide a more complete evaluation.

As illustrated in Fig. \ref{fig:styleTransResult}G, we visualise the comparison results in a two-dimensional Cartesian coordinate system, where the x-axis represents $\mathrm{FID}(\cdot \| \mathcal{V})$ and the y-axis represents $\mathrm{WD}(\cdot \| \mathcal{V})$. Each image set is mapped to a point in this space. The origin corresponds to the experimental dataset $\mathcal{V}$, representing zero style gap. The point labelled $\mathcal{U}$ denotes the pure simulated dataset, and the Euclidean distance between $\mathcal{U}$ and the origin captures the initial style gap between simulation and experiment, shown as the shaded blue region. The style-translated and perturbed versions of $\mathcal{U}$ are plotted in the same coordinate space. The ideal translation trajectory, where a simulated image set is perfectly converted into the experimental style, is shown by the dashed arrow. The solid arrow from $\mathcal{U}$ to $\mathcal{\tilde{V}}$ illustrates the style translation of generator $G_\mathcal{U}$. This transformation largely reduces both the FID and Wasserstein distances, effectively reducing the style gap between the original (blue) region and a smaller (orange) region, demonstrating the success of the learned translation. In contrast, the points corresponding to handcrafted permutations show limited or no reduction in style gap. Notably, adding salt-and-pepper noise actually increases the distance from the experimental distribution, moving the simulated images further away from the target domain.  For more information on the details of WD and FID, please refer to the section on Materials and Methods.

In summary, the CycleGAN-based forward translator $G_\mathcal{U}$ provides an effective solution for bridging the style gap between simulated and experimental AFM images. Unlike the handcrafted perturbations, which may unintentionally move the data distribution further from the experimental style, the learned translation yields significantly improved realism and authenticity, as supported by both qualitative visual comparisons and quantitative distribution metrics.

\section{Structure discovery model training and predictions}\label{sec:structureDiscovery}
\begin{figure*}[ht!]
    \centering
    \includegraphics[width=\textwidth]{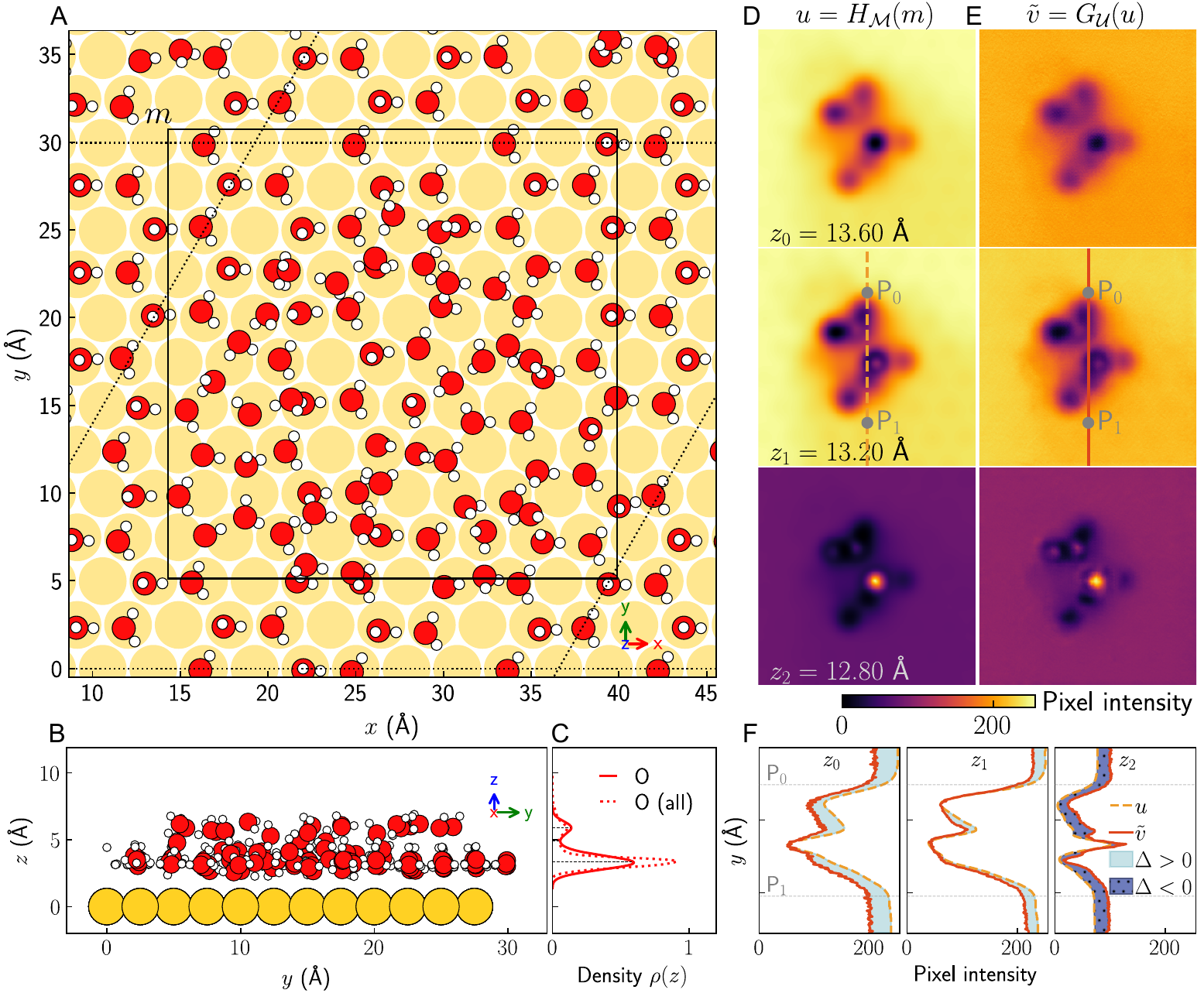}
    \caption{\textbf{Schematic illustration of a training sample.} (\textbf{A} to \textbf{B}) The simulated atomic configuration $m$ of a bi-layer water molecule cluster on the Au(111) surface from top and side views, respectively. (\textbf{C}) The probability density $\rho(z)$ of the oxygen atoms along the $z$ axis, where solid and dashed lines correspond to the density of this specific configuration $m$ and the mean density of many configurations $\mathcal{M}$. (\textbf{D}) The simulated AFM image $u$ of the configuration $m$ at different heights from $z_0$ to $z_2$ through the PPM $H_\mathcal{M}$. (\textbf{E}) The experimental style AFM image $\tilde{v}$ obtained by a simulation-to-experiment style translation model $G_\mathcal{U}$ with the simulated AFM image $u$ as the input. (\textbf{F}) The pixel intensity comparisons between simulated $u$ and experimental style $\tilde{v}$ AFM images along the direction from P$_0$ to P$_1$ at different heights, where the shadow areas indicate the difference of pixel intensity between experimental style $\tilde{v}$ and simulated $u$ AFM images.}
    \label{fig:dataset}
\end{figure*}

\noindent\textbf{Two types of training data for structure discovery model.} With the availability of experimental-style AFM images generated by the forward generator, we construct training datasets for the structure discovery model that maps 3D AFM images to atomic configurations. In this study, we use bilayer water configurations $\mathcal{M}$ absorbed on the Au(111) surface, obtained from simulations based on DFT and machine learning potentials \cite{Priante2024}. Figure \ref{fig:dataset}A demonstrates one example of an atomic configuration $m$, where the dashed lines indicate the unit cell and the rectangle indicates the $xy$ region used for AFM image simulations via PPM $H_\mathcal{M}$.  A side view of this configuration is shown in Fig. \ref{fig:dataset}B. The corresponding oxygen density profiles along the $z$ axis are plotted in Fig. \ref{fig:dataset}C. The solid line represents the density of the single configuration $m$, while the dashed line shows the average density across the configuration set $\mathcal{M}$. Two distinct peaks around $z = 3.3$ \AA{}  and $z = 5.9$ \AA{} reveal the bilayer nature of the structures. Figure \ref{fig:dataset}D shows three representative slices from a 3D simulated AFM image $u$, generated from configuration $m$. Pixel intensities reflect the frequency shift signals in AFM imaging.  To create the corresponding experimental-style AFM image $\tilde{v}$, we apply the trained forward style generator $G_\mathcal{U}$ to each 2D slice of $u$ and stack them to reconstruct the 3D AFM image, as shown in Fig. \ref{fig:dataset}E. Figure \ref{fig:dataset}F shows the detailed difference at the line P$_0$P$_1$ between simulated and experimental style AFM images at different heights. We can see that $\tilde{v}$ is noisier and contains more details compared to the original simulated image $u$. The pixel-wise differences $\Delta = \tilde{v} - u$ are shown as the shaded regions in Fig. \ref{fig:dataset}F, indicating the style change. 

To examine whether decreasing the style gap in training data helps to improve structure prediction on experimental AFM images, we design two kinds of training datasets: 1. Simulation dataset: Each training sample $(u, m)$ contains a 3D simulated AFM image $u$ and its corresponding atomic configuration $m$. 2. Experimental-style dataset: Each training sample $(\tilde{v}, m)$ contains a 3D experimental-style AFM image $\tilde{v}$ and the same atomic configuration $m$. Through the style translation, we transform the simulated data distribution into a new experimental-like distribution that approximates the real experimental distribution. It follows logically that a model trained on this distribution, denoted $F_\mathcal{\tilde{V}}$, should generalise better to real experimental AFM images than a model $F_\mathcal{U}$ trained solely on the simulation dataset. In the following discussion, we evaluate and compare the performance of models trained on different datasets to test our hypothesis and validate the impact of style translation on the accuracy of structure discovery. For more details on simulation data generation, including DFT calculations and AFM simulations, and the architecture of the structure discovery model, please refer to our previous studies \cite{Oinonen2022, Priante2024}.

\noindent\textbf{Atomic structure prediction on experimental AFM images.}
\begin{figure*}[ht!]
    \centering
    \includegraphics[width=\textwidth]{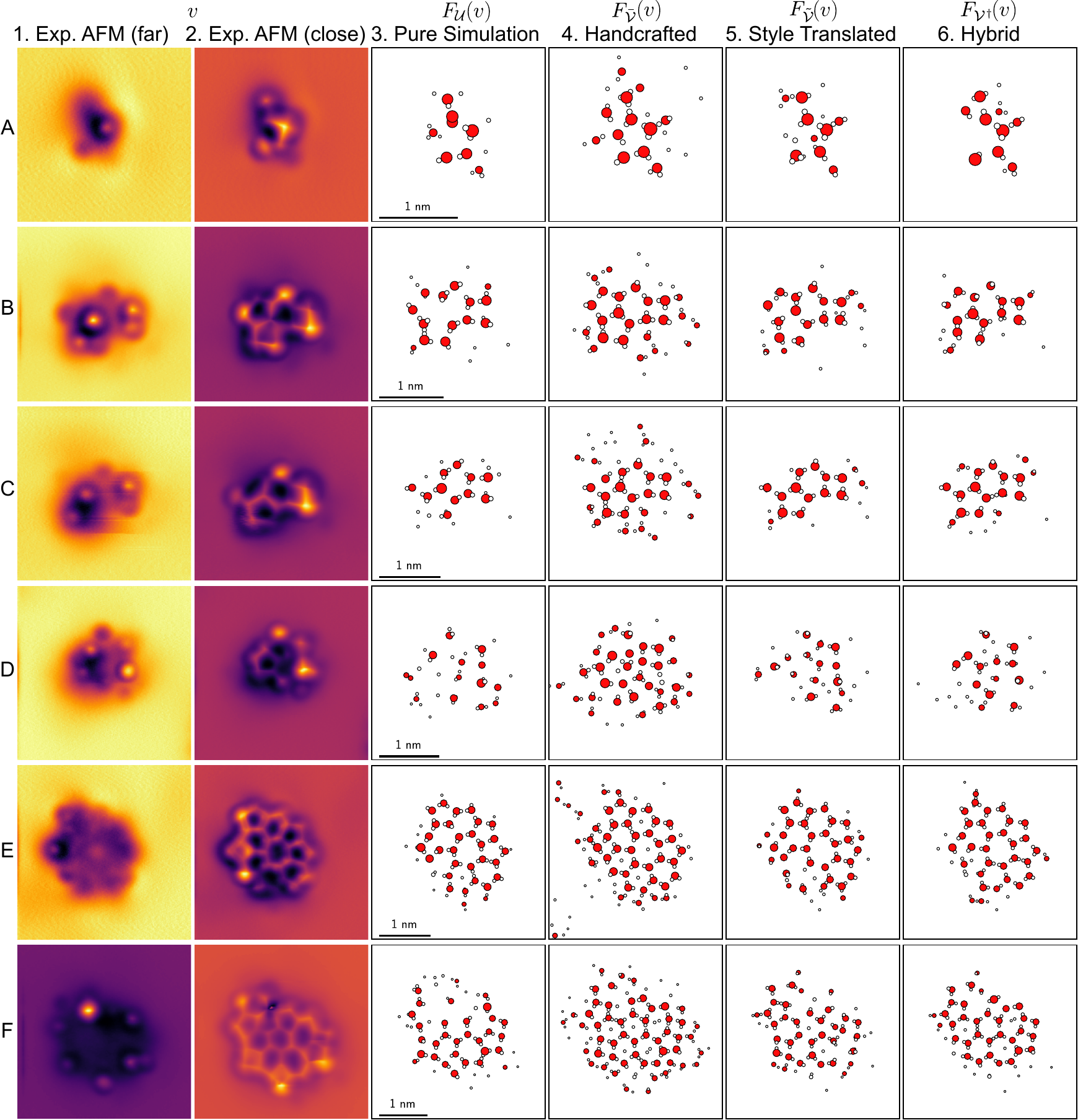}
    \caption{\textbf{Atomic configuration predictions from experimental AFM images using structure discovery models trained on different datasets.} Models trained on four types of images, including pure simulated AFM images $\mathcal{U}$, handcrafted perturbed images $\bar{\mathcal{V}}$, style translated images $\tilde{\mathcal{V}}$, and hybrid images $\mathcal{V}^{\dagger}$ where both style translation and handcrafted perturbation are applied.}
    \label{fig:v0v1_predictions}
\end{figure*}
Figure \ref{fig:v0v1_predictions} presents the predicted atomic configurations from different structure discovery models. Each model $F$ receives an input of a 3D experimental AFM image with vertical resolution of $\Delta z = 0.1$ \AA{} along the surface normal. The first two columns display representative 2D AFM slices of the input 3D AFM image at relatively far and close distances to the sample surface, respectively. Column 3 shows the atomic configuration predictions from the model $F_\mathcal{U}$, which is trained solely on the simulation dataset. Column 4 shows predictions from model $F_{\bar{\mathcal{V}}}$ trained on the dataset with handcrafted perturbations. Column 5 shows the results from $F_{\mathcal{\tilde{V}}}$, trained on experimental-style images generated via style translation. Column 6 presents predictions from $F_{\mathcal{V}^{\dagger}}$, trained on hybrid images combining style translation and handcrafted perturbations. For the style translation, we use cycle-consistency and identity loss weights of $\lambda_\mathrm{c}=20$ and $\lambda_\mathrm{i}=1$, respectively. The sizes of the atoms in each prediction reveal the relative heights of each atom in the $z$-axis, allowing for a visual assessment of vertical structure. Rows from A–F represent different experimental AFM samples. 

Qualitatively, model $F_\mathcal{U}$ predicts fewer atoms compared to the other models. This raises a key question: are these atoms genuinely absent in the experiment, or are they simply missed by the model due to limitations in generalising from simulation data? To investigate this, we run PPM simulations using the predicted atomic configurations and compare the resulting AFM images with the experimental inputs. A good prediction should yield a recovered simulation that closely matches the experimental image. The corresponding results are shown in Figs.   \ref{fig:recoveredPPAFM_pure}–\ref{fig:recoveredPPAFM_Hybrid} in Supporting Information (SI). These comparisons indicate that $F_\mathcal{U}$ indeed misses atoms, particularly evident in sample F (Fig. \ref{fig:recoveredPPAFM_pure}). Among the other models, predictions from $F_\mathcal{\tilde{V}}$ and $F_{\mathcal{V}^{\dagger}}$ tend to be more conservative than those from $F_{\bar{\mathcal{V}}}$, especially for the atoms positioned at lower $z$ values. However, the recovered AFM images from all three models appear similar (Figs. \ref{fig:recoveredPPAFM_handcrafted}–\ref{fig:recoveredPPAFM_Hybrid}), suggesting that the suppressed underlying atoms have limited influence on the overall AFM signal. Notably, the experimental-style-trained models exhibit enhanced robustness against image noise, suggesting improved generalisation to experimental data features. 

However, since the exact atomic structures from these AFM images are unknown, we cannot perform a direct accuracy evaluation like that on the simulation data. Consequently, visual inspection alone is insufficient to assess model performance. Here, we introduce an evaluation approach based on local structural properties that bypasses the need for ground-truth atomic configurations.

\section{Performance evaluations}\label{sec:performanceEva}
\noindent{\bf Structural metrics.}
Evaluating model performance on experimental AFM images is inherently challenging due to the absence of ground-truth atomic configurations, unlike in simulation datasets, where both AFM images and corresponding atomic structures are known. In contrast to typical machine learning tasks, such as image classification, where humans can often establish ground truth, structure discovery presents a fundamentally harder problem: even experts cannot reliably determine the exact atomic configurations from AFM images alone.  Nevertheless, even without exact atomic labels, predicted structures should still exhibit physically meaningful local properties, such as atom-atom distance and angle distributions, that are consistent with those obtained from first-principles simulations such as DFT \cite{Hong2024}. Therefore, instead of comparing each prediction to an unavailable ground truth, we assess model performance by comparing the statistical distributions of local structural properties across many predicted configurations. This provides an indirect yet meaningful evaluation of the physical validity of the predicted structures. 

\begin{figure*}[ht!]
	\centering
	\includegraphics[width=\textwidth]{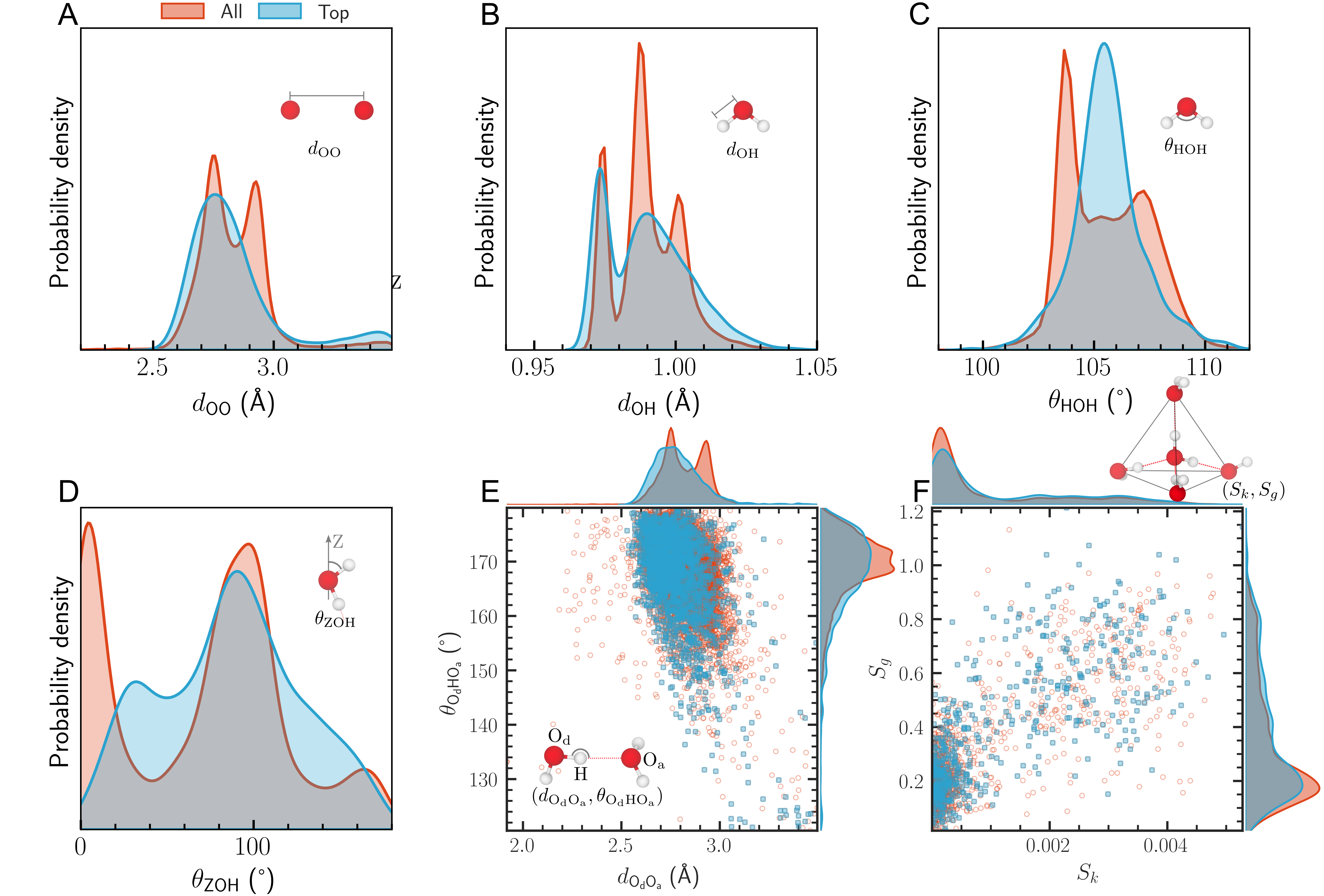}
	\caption{\textbf{Local structural distributions of water molecules in the configuration set $\mathcal{M}$ and in the top interfacial layer obtained from simulations.} (\textbf{A}) Distribution of oxygen–oxygen distances $d_\text{OO}$ within a cutoff of 3.5 \AA{}. (\textbf{B}) Distribution of oxygen–hydrogen distances $d_\text{OH}$ within a cutoff of 1.25 \AA{}. (\textbf{C}) Distribution of the intramolecular bond angle $\theta_\text{HOH}$. (\textbf{D}) Distribution of the angle $\theta_\text{ZOH}$ between the free OH bond and the surface normal ($z$ direction). (\textbf{E}) Joint distribution of the donor–acceptor oxygen distance $d_{\text{O}\text{d}\text{O}\text{a}}$ and the hydrogen bond angle $\theta_{\text{O}\text{d}\text{H}\text{O}\text{a}}$, used to identify hydrogen bonds based on geometric criteria ($d_{\text{O}\text{d}\text{O}\text{a}} < 3.5$ \AA{} and $\theta_{\text{O}\text{d}\text{H}\text{O}\text{a}} > 120^\circ$).
(\textbf{F}) Joint distribution of translational ($S_k$) and orientational ($S_g$) tetrahedral order parameters, which together characterise the local structural order.}
	\label{fig:theory}
\end{figure*}
\noindent\textbf{Structural property distributions.}
As shown in Fig. \ref{fig:theory}, we calculated local structural distributions for two cases: (1) all the water molecules in the configuration set $\mathcal{M}$, which includes both the top and bottom layers on the Au(111) surface, and (2) only the top layer water molecules. This comparison allows us to see the differences between the two layers. 

Figure \ref{fig:theory}A demonstrates the oxygen-oxygen distance $d_\text{OO}$ distribution within a $3.5$ \AA{} cutoff, characterizing intermolecular structure. A dominant peak appears around 2.75 \AA{} for both cases, while an additional peak near $2.95$ \AA{} arises from the interaction between the bottom layer of water molecules and the Au(111) surface.  Figure \ref{fig:theory}B and C shows the distributions of oxygen-hydrogen distance $d_\text{OH}$ and the angle $\theta_\text{HOH}$, respectively, both within a  $1.25$ \AA{} cutoff to capture the intramolecular geometry. Structural differences are observed for the top layer, reflecting its greater configurational freedom compared to the bottom layer, which is more constrained by the substrate. We use free OH bonds \cite{Shen2006, Tang2017, Fujie2024} to capture the surface-related orientation of water molecules.  Figure \ref{fig:theory}D demonstrates the distribution of the angle $\theta_\text{ZOH}$ between the free OH bond vector and the surface normal. To analyse hydrogen bonding, we calculate the joint distribution of donor-acceptor oxygen distance $d_{\text{O}_\text{d}\text{O}_\text{a}}$ and hydrogen bond angle $\theta_{\text{O}_\text{d}\text{H}\text{O}_\text{a}}$, shown in Fig. \ref{fig:theory}E. A hydrogen bond is considered present if  $d_{\text{O}_\text{d}\text{O}_\text{a}} < 3.5$ and $\theta_{\text{O}_\text{d}\text{H}\text{O}_\text{a}} > 120^{\circ}$. Each point in the space corresponds to a donor–acceptor pair that satisfies this geometric criterion. In addition, inspired by studies on using order parameters to study the local water structures \cite{OffeiDanso2022, Donkor2023}, to evaluate the structure order within the water network, we calculate the joint distribution of two tetrahedral order parameters: the translational order parameter $S_k$ and the orientational tetrahedral order parameter $S_g$ is also calculated \cite{CHAU1998, DubouDijon2015}. The translational tetrahedral order $S_k$ is defined as 
\begin{equation}
S_k = \frac{1}{3} \sum_{k=1}^4 \frac{\left(r_k - \bar{r}\right)^2}{4\bar{r}^2},
\end{equation}
where $r_k$ is the radial distance from the central oxygen atom to the $k$-th peripheral oxygen atom, and $\bar{r}$ is the arithmetic mean of the four radial distances. This parameter quantifies the variance in radial distances, with $S_k = 0$ for a perfect tetrahedron and increasing as the structure becomes distorted. The orientational tetrahedral order $S_g$ is defined as 
\begin{equation}
S_g = \frac{3}{8} \sum_{j=1}^3 \sum_{k=j+1}^4 \left( \cos \psi_{j,k} + \frac{1}{3} \right)^2
\end{equation}
where $\psi_{j, k}$ is the angle between bonds $j$ and $k$ at the central oxygen. A perfect tetrahedron yields $S_g=0$, while random angular arrangements yield an average $\langle S_g \rangle \approx 1$ due to the normalisation factor. The joint distribution of $S_k$ and $S_g$ is illustrated in Fig. \ref{fig:theory}F, where each point corresponds to a local environment consisting of a central oxygen atom and its four nearest neighbours within a 3.5 \AA{} cutoff.

\begin{figure*}[ht!]
	\centering
	\includegraphics[width=\textwidth]
        {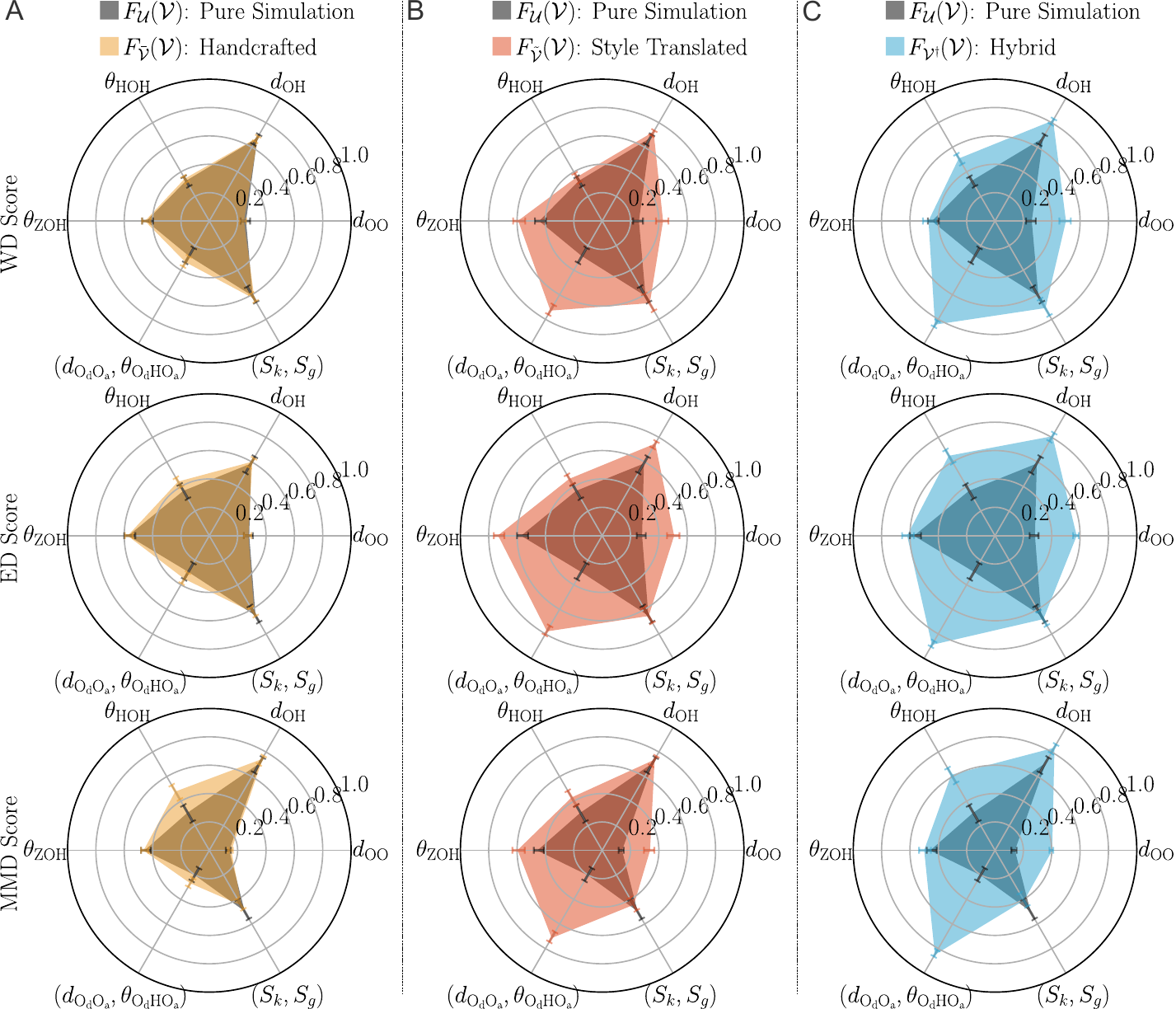}
	\caption{\textbf{Performance evaluations based on the structural properties,  using three distance metrics: Wasserstein distance, energy distance, and maximum mean discrepancy.} Panels (A), (B), and (C) show comparisons of models $F_{\bar{\mathcal{V}}}$, $F_{\tilde{{\mathcal{V}}}}$ and $F_{{\mathcal{V}}^{\dagger}}$, respectively, each evaluated against the baseline model $F_{\mathcal{U}}$. Distances are computed between predicted distributions and reference theoretical distributions derived from top-layer water molecules.}
	\label{fig:Radar_top}
\end{figure*}

\noindent\textbf{Performance evaluations based on distributional distances.}
Figure \ref{fig:Radar_top} shows a comparative performance evaluation of three structure discovery models trained on modified datasets: $F_{\bar{\mathcal{V}}}$ (handcrafted perturbations), $F_{\tilde{{\mathcal{V}}}}$ (style-translated) and $F_{{\mathcal{V}}^{\dagger}}$ (hybrid),  against the baseline model $F_\mathcal{U}$ that is trained on pure simulated AFM images. Each row corresponds to one of three distributional distance metrics used to assess structural fidelity: Wasserstein distance (WD, top row), energy distance (ED, middle row), and maximum mean discrepancy (MMD, bottom row). For detailed calculations of these metrics, please refer to the Materials and Methods section.

Each radar chart visualises normalised performance scores across six structural properties, as previously defined in Fig. \ref{fig:theory}: the oxygen–oxygen distance $d_\text{OO}$, the oxygen–hydrogen distance $d_\text{OH}$, the angle $\theta_\text{HOH}$, the free OH orientation angle $\theta_\text{ZOH}$, the hydrogen bond geometry ($d_{\text{O}\text{d}\text{O}\text{a}}, \theta_{\text{O}\text{d}\text{H}\text{O}\text{a}}$), and the tetrahedral order parameters ($S_k$, $S_g$). For each property, we compute the distance between the predicted and theoretical distributions. These distances are normalised using min-max normalisation across all models in our computational experiments with varying hyperparameter settings ($\lambda_\mathrm{c}$, $\lambda_\mathrm{i}$). To convert distances into performance scores, we use $1 - \text{normalised distance}$, such that higher values indicate better agreement with the theoretical distributions. The reference theoretical target distributions in Fig. \ref{fig:Radar_top} are obtained from the top-layer water molecules in the configuration set $\mathcal{M}$ as shown in Fig. \ref{fig:theory}. In each plot, the gray polygon represents the performance of the baseline model $F_\mathcal{U}$, and the orange, red, and blue polygons represent the performance of models trained with handcrafted, style-translated, and hybrid datasets, respectively. The error bars are obtained from the standard error of different replicas of the structure discovery model trained on the same dataset. For more details on distribution comparisons for these six structural properties, please refer to Figs. \ref{fig:OO_compare}–\ref{fig:OrderP_compare} in the SI. 

Figure \ref{fig:Radar_top}A shows that handcrafted perturbations provide moderate performance improvements in specific properties like angle $\theta_\text{HOH}$ and hydrogen bond geometry ($d_{\text{O}\text{d}\text{O}\text{a}}, \theta_{\text{O}\text{d}\text{H}\text{O}\text{a}}$), but offer little benefit for other structural metrics. Figure \ref{fig:Radar_top}B represents that style translation yields broad improvements across most structural properties, particularly in oxygen-oxygen distance and hydrogen bond geometry ($d_{\text{O}\text{d}\text{O}\text{a}}, \theta_{\text{O}\text{d}\text{H}\text{O}\text{a}}$). However, performance gains are less evident for order parameters ($S_k$, $S_g$). The tetrahedral order parameters are inherently less local than the others, requiring the central oxygen atom to have another four oxygen neighbours within 3.5 \AA{}, posing challenges in accurately computing the distributional distances when limited tetrahedral environments are found in the predicted configurations (see Fig. \ref{fig:OrderP_compare} in SI).  Figure \ref{fig:Radar_top}C shows that hybrid datasets achieve the most balanced and consistent performance, improving nearly all metrics simultaneously. 
These results support the idea that reducing the image style gap between the simulated and experimental AFM images improves the accuracy and physical consistency of the predicted atomic structures.


\section{Conclusions}
With the goal of solving the inverse structure discovery problem of mapping real experimental AFM images to atomic configuration, we aim to fill the performance gap between models trained on simulated and experimental data in the challenging scenario where the ground-truth atomic configurations are unavailable. To bridge this gap,  we propose a data-driven,  unpaired image-to-image style translation approach that significantly reduces the style discrepancy between simulated and experimental AFM images. Our approach relies only on unpaired samples from simulation and experiment domains, making it particularly well-suited for real-world scenarios where such paired data is impractical. 

We demonstrate the effectiveness of this method using the example of water structure discovery on Au(111). By replacing the unavailable experimental training data with style-translated simulated AFM images, we show that structure discovery models can achieve significantly better performance on real experimental inputs. The style-translated dataset exposes the model to more realistic conditions, helping it focus on essential atomic features rather than noise and artefacts. These results support our hypothesis that reducing the style gap improves model performance on experimental AFM images. In addition, we propose an evaluation approach based on physically meaningful structural properties to address the lack of ground-truth atomic structures for experimental AFM images.

Overall, our work offers a practical pathway toward closing the simulation-to-experiment gap in AFM structure discovery. In this study, structural properties are used solely for evaluation purposes, but they are not used to guide model training. This raises an important question: can such structural properties be integrated as constraints during the model training to discourage physically implausible predictions? Another challenge lies in balancing robustness and sensitivity. While generalising across noisy experimental conditions improves stability, it may also bring the sensitivity loss to subtle atomic features. Understanding and balancing this trade-off is essential for developing high-fidelity models. Finally, we envision future structure discovery frameworks that can provide confidence scores for each predicted atom, helping us assess the trustworthiness of the model's outputs.

\section{Materials and Methods}

\noindent\textbf{Wasserstein distance.} Wasserstein distance \cite{arjovsky2017wgan, weng2017gan} (also called Earth Mover's distance) is a measure of the dissimilarity between two distributions. Intuitively,  it quantifies the minimum ``effort" required to transform one distribution into another, where the effort is measured by the amount of probability mass that must be transported and the distance it must be moved. 
Mathematically, the Wasserstein distance between two distributions $X$, $Y$ is defined as follows:
\begin{equation}
\begin{split}
        \text{WD}(X, Y)  &= \inf_{\pi \in \Pi(X, Y)} \int_{\mathbb{R}\times\mathbb{R}} \|x - y\| \, \mathrm{d} \pi(x, y) \\
        &= \inf_{\pi \in \Pi(X, Y)} \mathbb{E}_{(x, y) \sim \pi} \big[ \| x - y \| \big].\\
\end{split}
\end{equation}
Here,  $\Pi(X, Y)$ denotes the set of all joint distributions $\pi(x, y)$ whose 
marginals are respectively $X$ and $Y$. The joint distribution $\pi(x, y)$ specifies a transport plan indicating how much mass should be transported from $x$ to $y$. The Wasserstein distance is then the minimum cost of the transport plan. 

\noindent\textbf{Fréchet inception distance}. Fréchet inception distance (FID) was first introduced to evaluate the performance of GAN models. The FID between two image distributions $X$ and $Y$ is computed as follows. Collect image samples $x_1, ..., x_m$, $y_1, ..., y_n$ from $X$, $Y$, respectively. Encode all samples $x_i$ and $y_i$ by computing the activations $\text{A}(x_i)$ and $\text{A}(y_i)$ of the final layer of the pretrained Inception network \cite{szegedy2015Inception}. Compute the sample means $\mu_1$, $\mu_2$ and the sample covariance matrices $\Sigma_1$, $\Sigma_2$ of the activations $\text{A}(x_i)$, $\text{A}(y_i)$. The FID is the WD between the two multivariate normal distributions $N(\mu_1, \Sigma_1)$ and $N(\mu_2, \Sigma_2)$ \cite{mathiasen2021BPFID}.

\begin{equation}
    \text{FID}(X, Y) = \left\| \mu_1 - \mu_2 \right\|_2^2 + \mathrm{tr}(\Sigma_1) + \mathrm{tr}(\Sigma_2) - 2 \cdot \mathrm{tr} \left( \sqrt{ \Sigma_1 \Sigma_2 } \right)
\end{equation}

\noindent\textbf{Energy distance.} Energy distance (ED) is a statistical distance used to measure the equality of distributions, whose name is derived from Newton's gravitational potential energy. The energy distance \cite{SZEKELY20131249} between the $d$-dimensional independent random variables $X$ and $Y$ is defined as
\begin{equation}
    \text{ED}(X, Y) = 2\mathbb{E}\|X - Y\|_d - \mathbb{E}\|X - X'\|_d - \mathbb{E}\|Y - Y'\|_d,
\end{equation}
where $X'$ is an independent and identically distributed (iid) copy of $X$, $Y'$ is an iid copy of $Y$, $\mathbb{E}$ is the expected value, and $\|\cdot\|$ is to denote Euclidean norm. 

\noindent\textbf{Maximum mean discrepancy.} Maximum mean discrepancy (MMD) is another distance measurement between random variables $X$ and $Y$, which is defined as the distance between their embeddings in the reproducing kernel Hilbert space (RKHS) \cite{JMLR_MMD}. MMD quantifies the dissimilarity between two distributions by comparing their mean representations in a high-dimensional feature space. Given two probability distributions $X$ and $Y$, the MMD between them is defined as follow: 
\begin{equation}
    \text{MMD}(X, Y) = \|\mu_X - \mu_Y\|_\mathcal{H},
\end{equation}
where $\mu_X$, $\mu_Y$ are the mean embeddings of $X$ and $Y$ in RKHS $\mathcal{H}$. We calculate the empirical estimation of MMD through
\begin{equation}
\begin{aligned}
    \text{MMD}^2(X, Y)  = & \left \|\frac{1}{n}\sum_{i=1}^n\varphi(x_i) - \frac{1}{m}\sum_{i=1}^m\varphi(y_i) \right \|_\mathcal{H}^2 \\
    = & \frac{1}{n^2} \sum_{i=1}^{n} \sum_{j=1}^{n} k(x_i, x_j)  + \frac{1}{m^2} \sum_{i=1}^{m} \sum_{j=1}^{m} k(y_i, y_j) \\
& - \frac{2}{nm} \sum_{i=1}^{n} \sum_{j=1}^{m} k(x_i, y_j),
\end{aligned}
\end{equation}
where kernel $k(x, y)$ is a function that measures the similarity between two data points $x$ and $y$. We use the Gaussian kernel: $k(x, y) = \exp \left ( - {\|x-y\|^2}/(2\sigma ^2)\right )$, where $\sigma$ is the bandwidth parameter. 

\noindent\section*{Data Availability}
The codes and training data used in this study will be made publicly available at these links upon publication of this work:
\href{https://github.com/SINGROUP/StyleTransAugment}{https://github.com/SINGROUP/StyleTransAugment},
\\
\href{https://doi.org/10.5281/zenodo.16828078}{https://doi.org/10.5281/zenodo.16828078}.

\section*{Conflict of interest}
The authors declare no competing financial interest.

\section*{Acknowledgments}
The authors thank Peter Liljeroth, Benjamin Alldritt, and Shuning Cai for their efforts in acquiring and providing the experimental images used for style translation training. J.H. thanks Johannes Haataja for inspiring discussions and Nan Cao for valuable suggestions on data visualisations. This work was supported by the World Premier International Research Center Initiative (WPI), MEXT, Japan, and by the Research Council of Finland (Projects 347319 and 346824). The authors acknowledge the computational resources provided by the Aalto Science-IT Project and CSC, Helsinki.

\bibliography{main.bib}
\clearpage 

\noindent{\bf \Large Supporting Information}
\vspace{0.5cm}
\newline

\noindent\textbf{CycleGAN network architectures and training details.} We use ResNet-based generators with 6 residual blocks to perform style translations. Each generator takes a single-channel (grayscale) input and outputs a single-channel image. The architecture consists of: (a) an initial 7$\times$7 convolutional layer followed by instance normalisation and ReLU activation; (b) two downsampling layers (3$\times$3 convolutions with stride 2); (c) six residual blocks, each containing two 3$\times$3 convolutions with instance normalisation and skip connections; (d) two upsampling layers using transposed convolutions; and (e) a final 7$\times$7 convolutional layer followed by a Tanh activation function. We set the number of base convolutional filters to 16. For each discriminator, we adopt a PatchGAN architecture implemented as a CNN with two layers. The input is a single-channel AFM image, and the base number of filters is set to 16. The discriminator consists of a series of 4$\times$4 convolutional layers with a stride of 2, followed by instance normalisation and LeakyReLU activation. This configuration allows the discriminator to operate on overlapping image patches, making local decisions about authenticity, which encourages the generator to produce realistic fine-grained textures. The final layer outputs a single-channel prediction map. We trained our CycleGAN model using PyTorch \cite{torch2019}. Training was carried out using greyscale AFM images with a size of 192$\times$192. We optimised the model using default settings in the CycleGAN framework, tuning key hyperparameters: the cycle-consistency loss weight and the identity loss weight. Training was carried out for 200 epochs. 

\begin{figure*}[ht!]
	\centering
	\includegraphics[width=\linewidth]{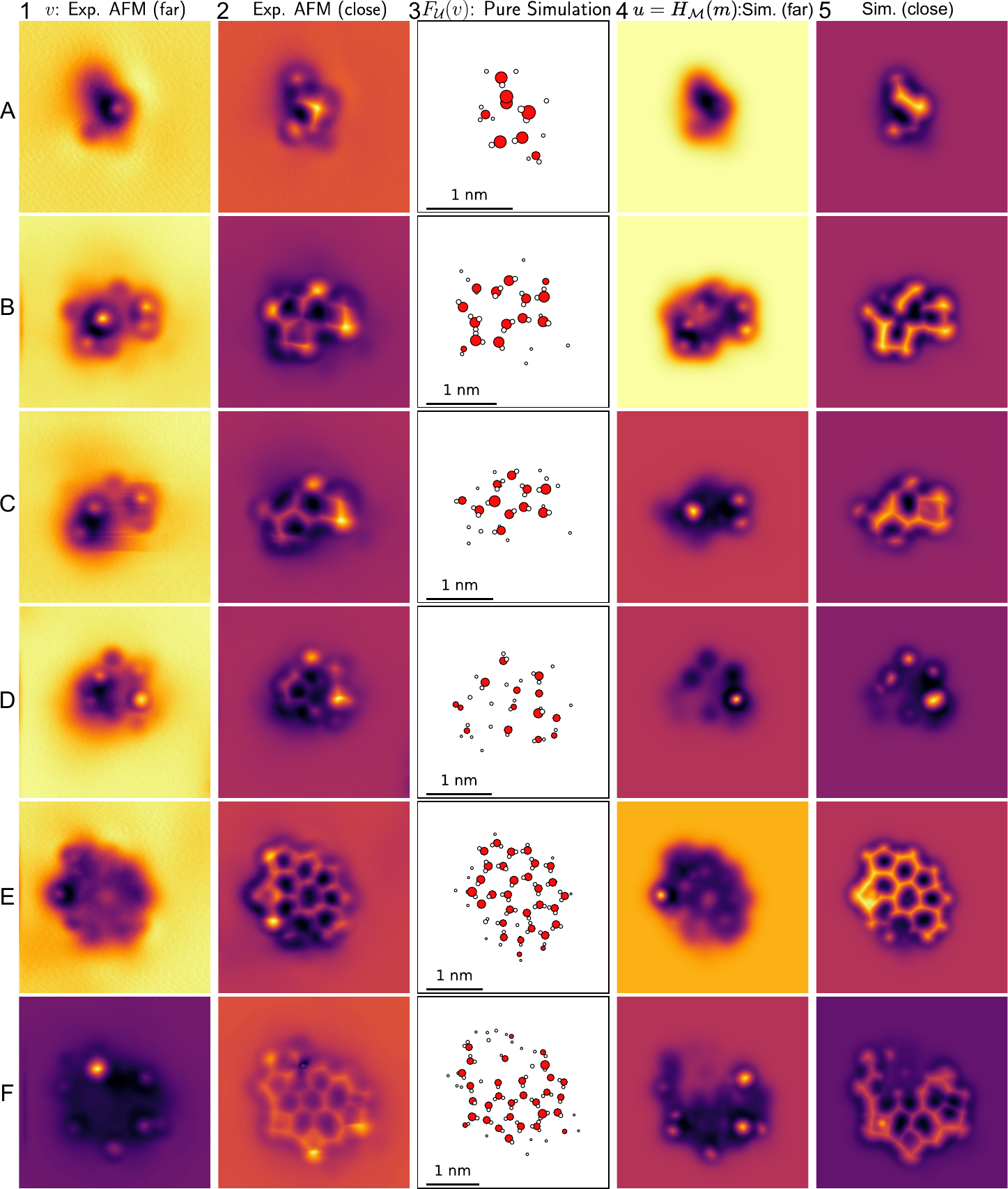}
	\caption{Structure predictions (Column 3), and corresponding PPM-simulated AFM images (Columns 4 and 5) for experimental AFM inputs, using model $F_\mathcal{U}$ trained on pure simulated AFM data.}
	\label{fig:recoveredPPAFM_pure}
\end{figure*}
\begin{figure*}[ht!]
	\centering
	\includegraphics[width=\linewidth]{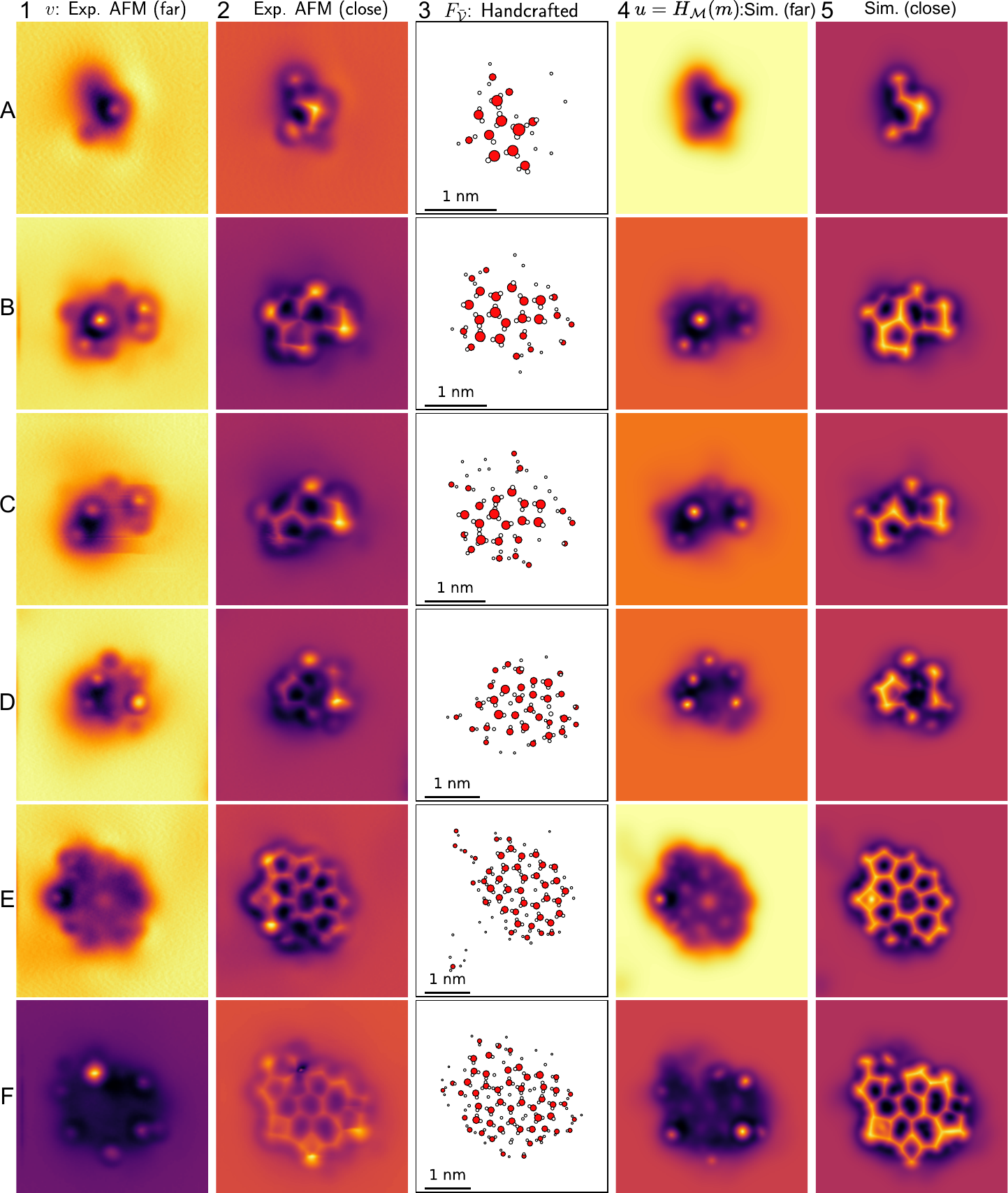}
	\caption{Structure predictions (Column 3), and corresponding PPM-simulated AFM images (Columns 4 and 5) for experimental AFM inputs, using model $F_\mathcal{\bar{V}}$ trained on images ${\mathcal{\bar{V}}}$ with handcrafted perturbations.}
\label{fig:recoveredPPAFM_handcrafted}
\end{figure*}
\begin{figure*}[ht!]
	\centering
	\includegraphics[width=\linewidth]{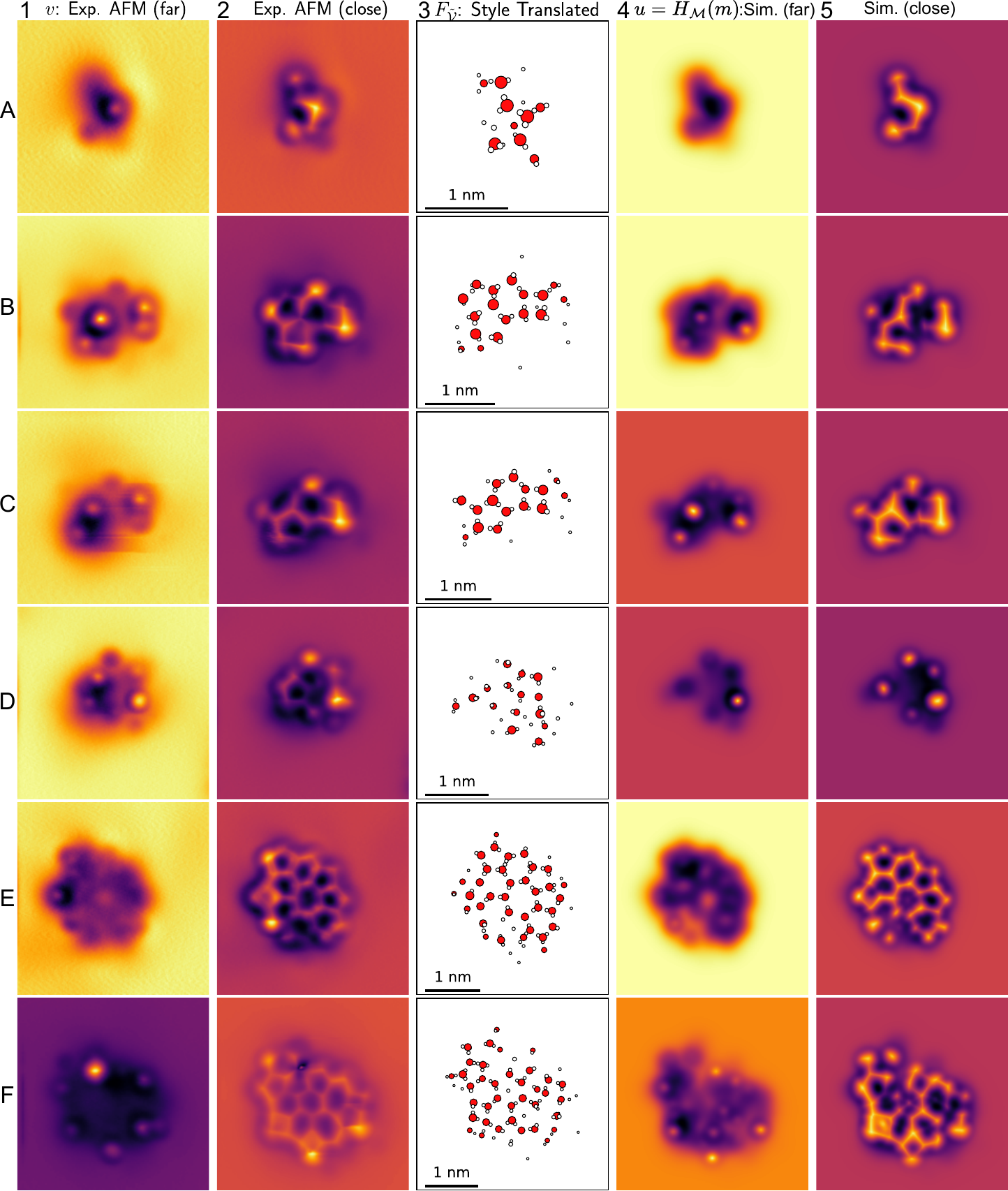}
	\caption{Structure predictions (Column 3), and corresponding PPM-simulated AFM images (Columns 4 and 5) for experimental AFM inputs, using model $F_\mathcal{\tilde{V}}$ trained on style-translated data with $\lambda_\mathrm{c} = 20, \lambda_\mathrm{i} = 1$.}
	\label{fig:recoveredPPAFM_StyleTransOnly}
\end{figure*}
\begin{figure*}[ht!]
	\centering
	\includegraphics[width=\linewidth]{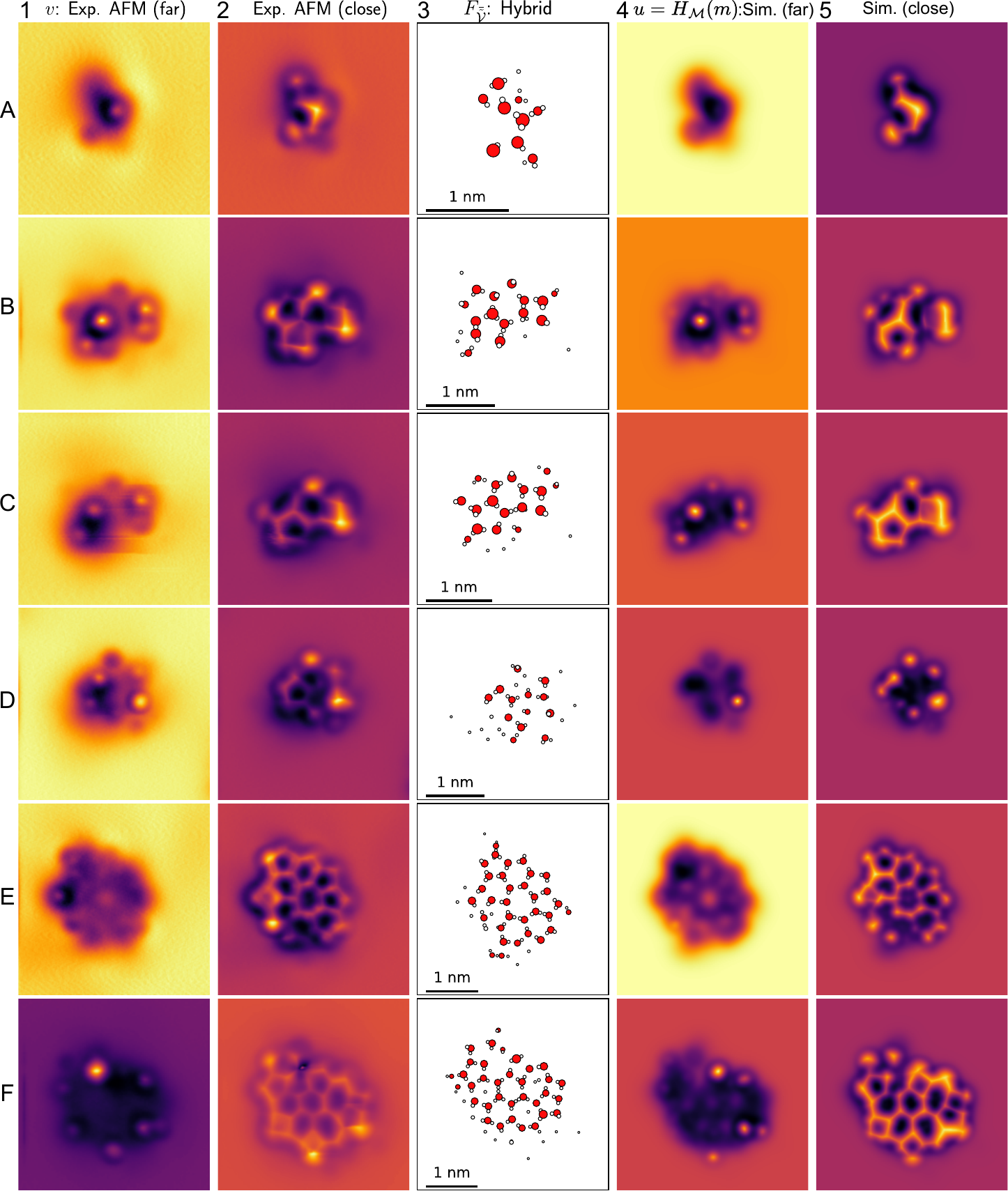}
	\caption{Structure predictions (Column 3), and corresponding PPM-simulated AFM images (Columns 4 and 5) for experimental AFM inputs, using model $F_{\mathcal{V^{\dagger}}}$ trained on images $\mathcal{V^{\dagger}}$ with hybrid modifications combined with handcrafted perturbations and style translations with $\lambda_\mathrm{c} = 20, \lambda_\mathrm{i} = 1$.}
	\label{fig:recoveredPPAFM_Hybrid}
\end{figure*}

\noindent{\bf AFM simulations for the predicted structures.}
Figures \ref{fig:recoveredPPAFM_pure} to \ref{fig:recoveredPPAFM_Hybrid} present the predicted atomic configurations and their corresponding PPM-simulated AFM images, obtained from four different structure discovery models.

\begin{figure*}[!ht]
	\centering
	\includegraphics[width=0.75\textwidth]{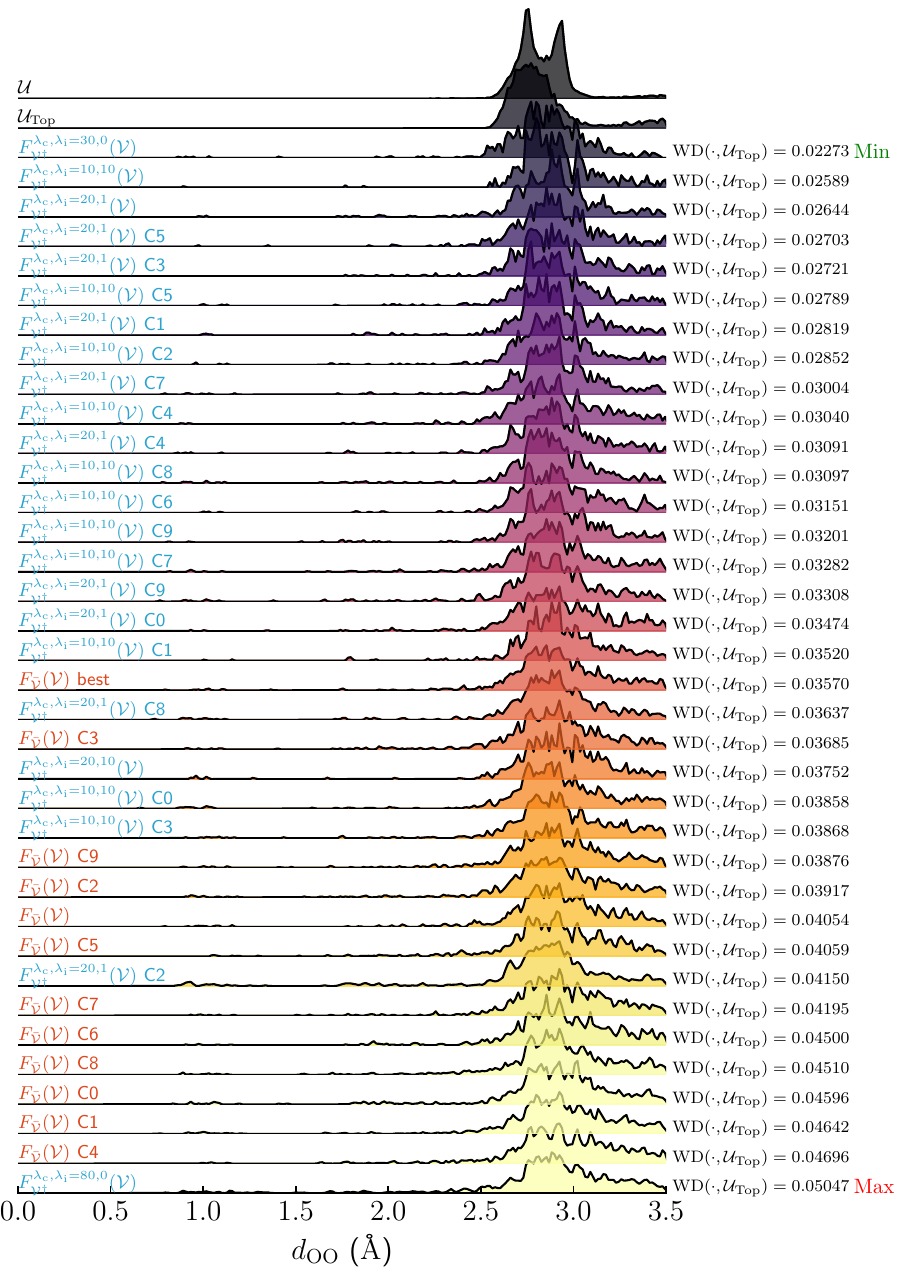}
	\caption{$d_\mathrm{OO}$ distributional comparisons between theoretical and predicted structures across different structure discovery models and datasets.}
	\label{fig:OO_compare}
\end{figure*}

\begin{figure*}[!ht]
	\centering
	\includegraphics[width=0.75\textwidth]{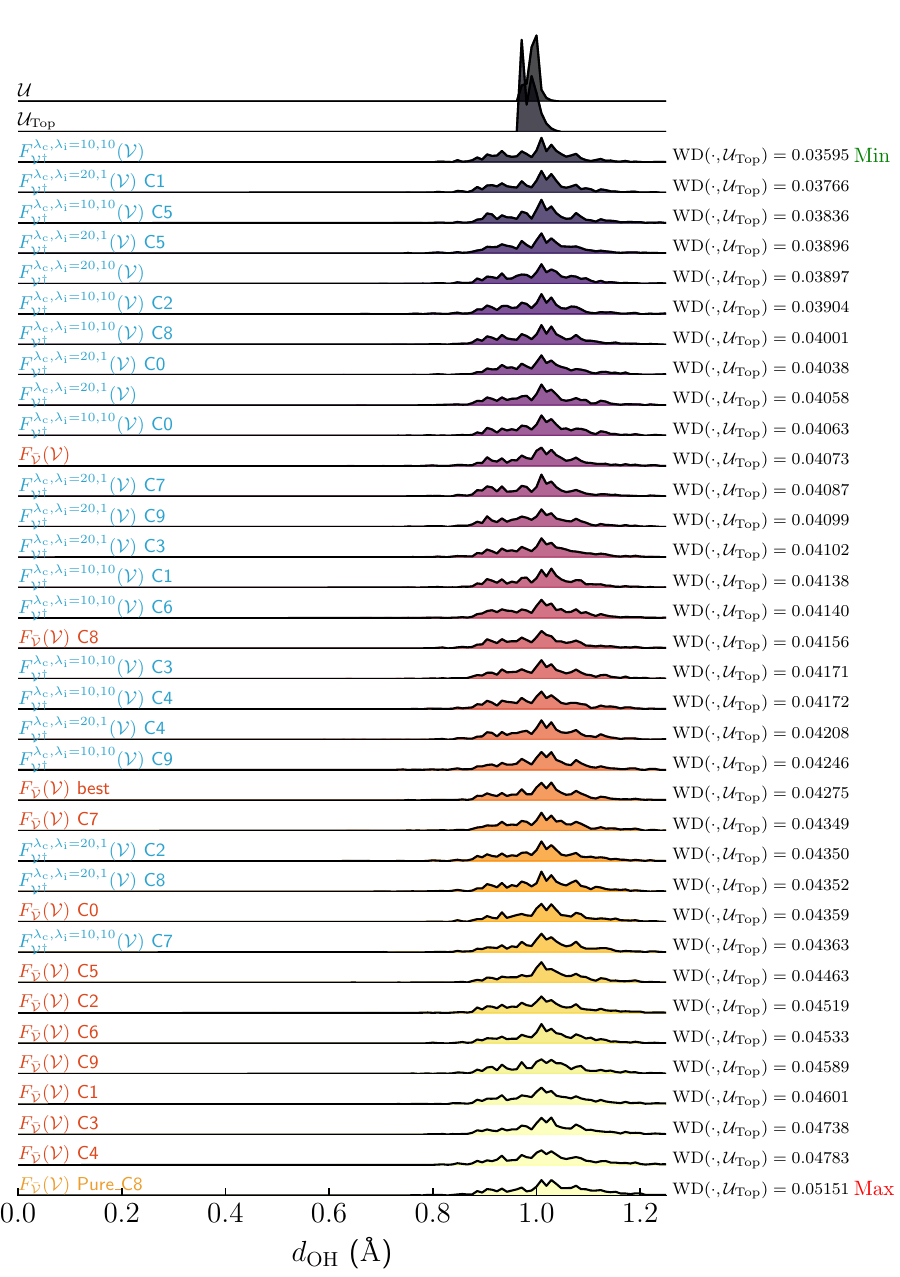}
	\caption{Distributional comparisons of $d_\mathrm{OH}$ between theoretical and predicted structures.}
	\label{fig:OH_compare}
\end{figure*}

\begin{figure*}[!ht]
	\centering
	\includegraphics[width=0.75\textwidth]{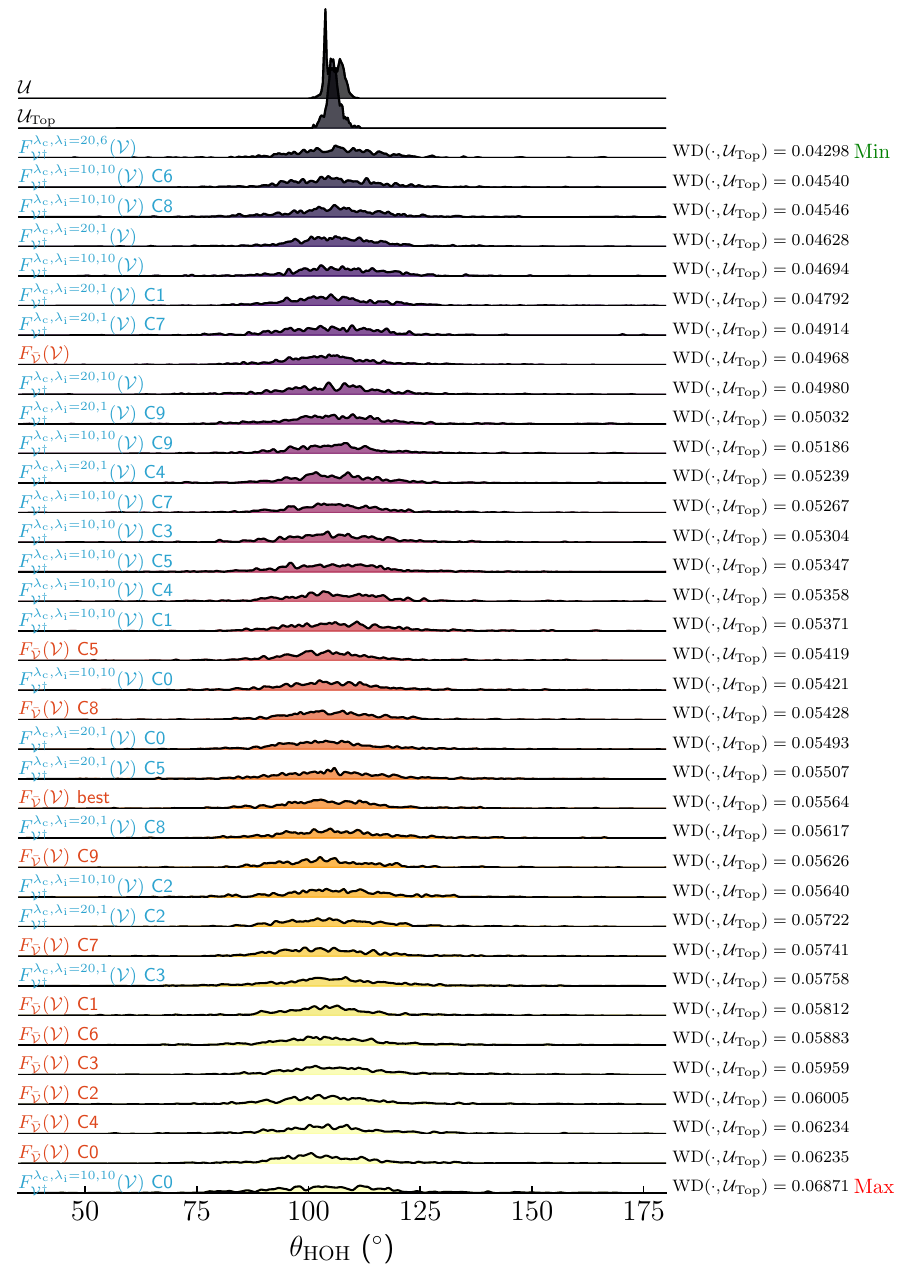}
	\caption{Distributional comparisons of $\theta_\mathrm{HOH}$ between theoretical and predicted structures.}
	\label{fig:HOH_compare}
\end{figure*}

\begin{figure*}
	\centering
	\includegraphics[width=0.75\textwidth]{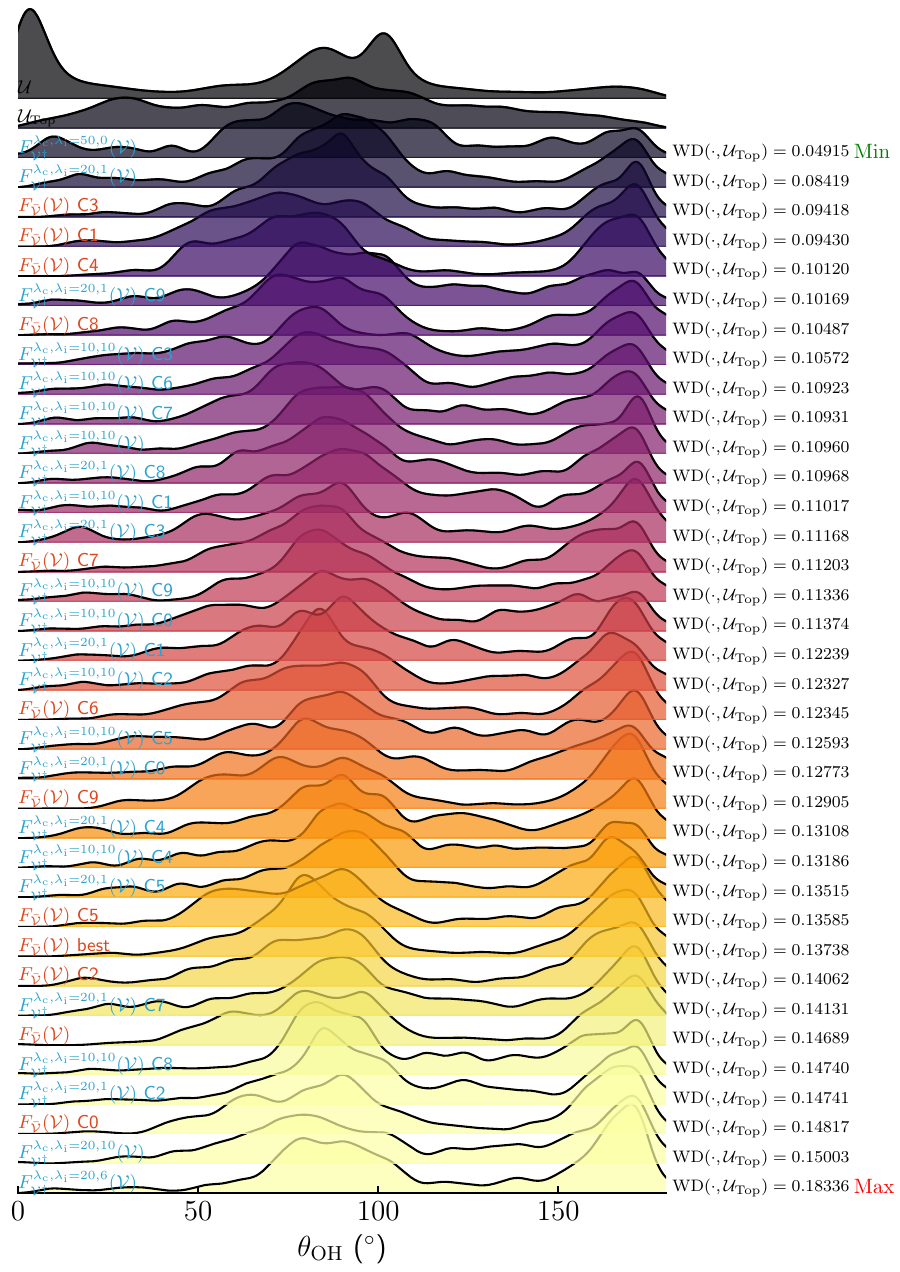}
	\caption{Distributional comparisons of the angle $\theta_\mathrm{ZOH}$ between the free OH bond and the surface normal ($z$ direction).}
	\label{fig:Theta_OH_compare}
\end{figure*}

\begin{figure*}[!ht]
	\centering
	\includegraphics[width=\textwidth]{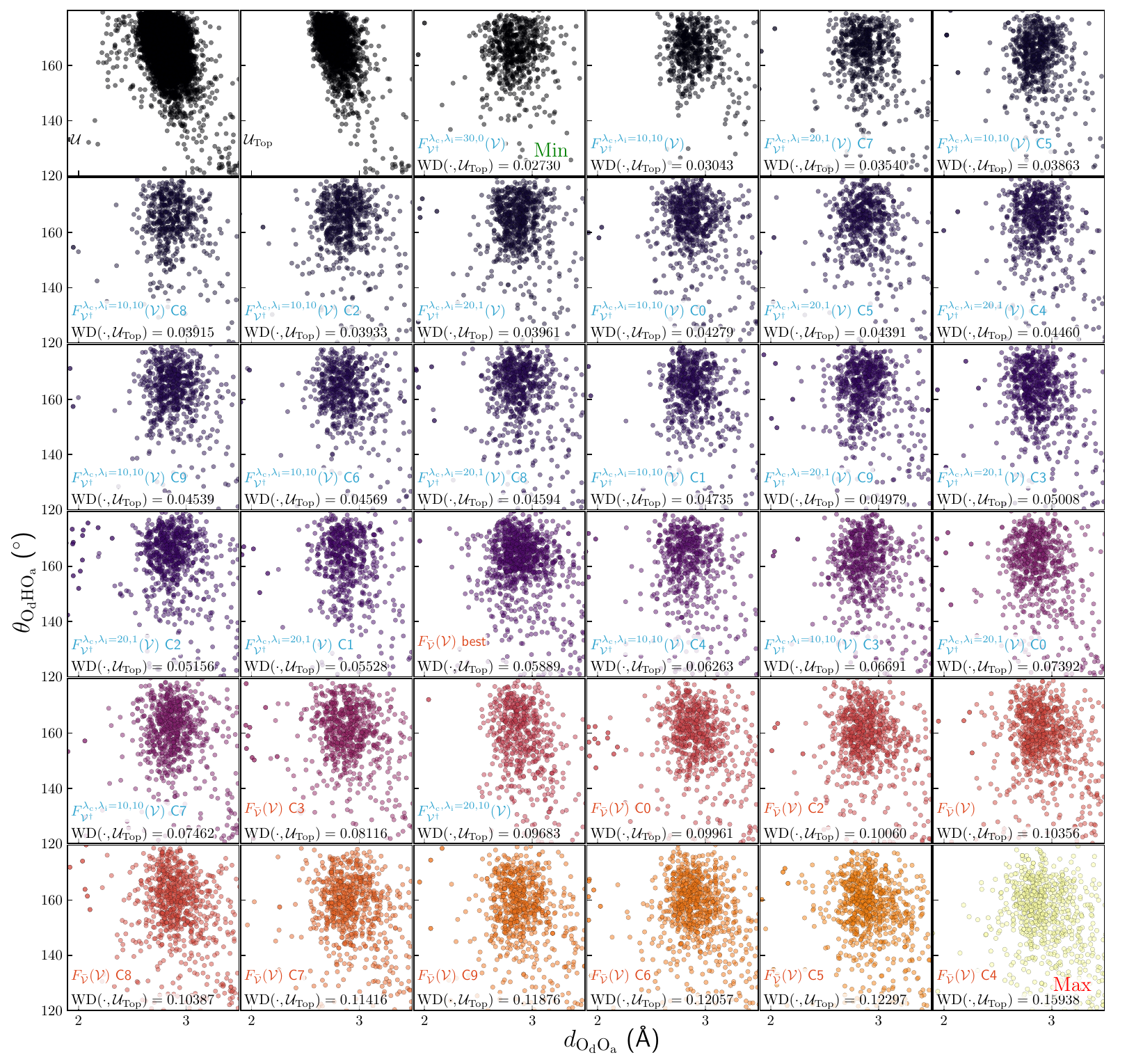}
	\caption{Comparisons of joint distributions of the donor–acceptor oxygen distance $d_{\text{O}_\text{d}\text{O}_\text{a}}$ and the hydrogen bond angle $\theta_{\text{O}_\text{d}\text{H}\text{O}_\text{a}}$.}
	\label{fig:Hbond_compare}
\end{figure*}

\begin{figure*}[!ht]
	\centering
	\includegraphics[width=\textwidth]{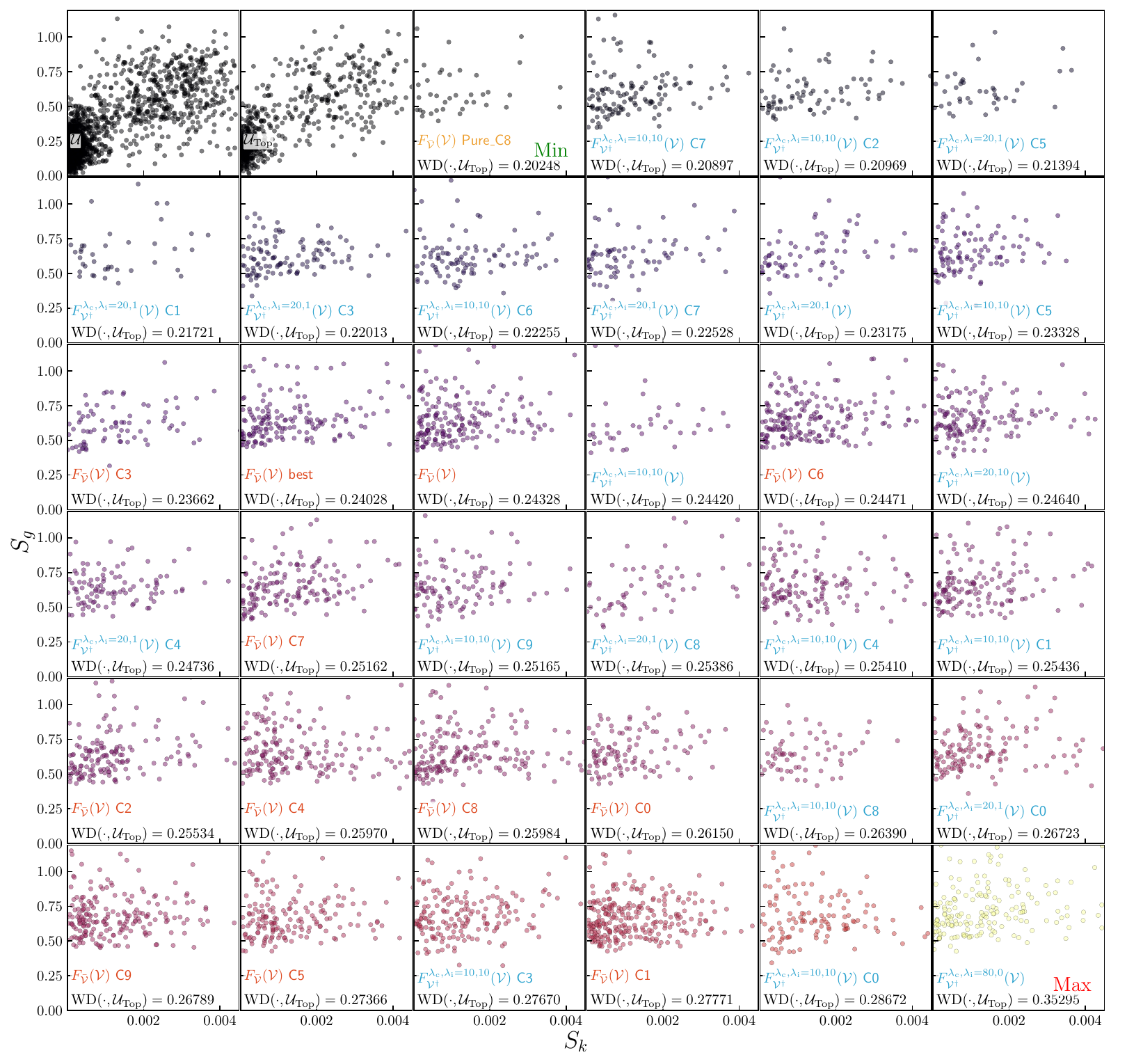}
	\caption{Comparisons of the joint distributions of translational ($S_k$) and orientational ($S_g$) tetrahedral order parameters.}
	\label{fig:OrderP_compare}
\end{figure*}

{\noindent\bf{Detailed distributional comparisons between theoretical and predicted structures.}}
Figures \ref{fig:OO_compare} through \ref{fig:OrderP_compare} show the detailed comparisons of six local structural property distributions between predicted configurations and reference structures of water molecules on Au(111). In these figures, the distributions of predicted structures are ordered based on the Wasserstein distance to the theoretical distribution of top-layer water molecules $\mathcal{U}_\mathrm{top}$, taking into account that the top layer of molecules is easier to predict by a structure discovery model compared to the underlying molecules. 

All structure discovery models share the same network architecture and training parameters. Two types of models are compared here: models $F_{\mathcal{\bar{V}}}$ trained on images $\bar{\mathcal{V}}$ with handcrafted perturbations and models $F_{\mathcal{V}^{\dagger}}$trained on hybrid modification images $\mathcal{V}^{\dagger}$ that combine with handcrafted perturbations and style translations. For models $F_{\mathcal{V}^{\dagger}}$, we selected and show two sets of modules with parameters $\lambda_{\mathrm{c}}, \lambda_{\mathrm{i}} = 10, 10$ and $\lambda_{\mathrm{c}}, \lambda_{\mathrm{i}} = 20, 1$. In addition, we also show the distributions of the smallest and largest distance $\mathrm{WD}(\cdot, \mathcal{U}_{\mathrm{top}})$ in all our computational experiments. Since training dynamics can lead to variability in model performance, we train 10 independent replicas per configuration. The results shown in Fig. \ref{fig:Radar_top} represent averaged performance over these replicas.

Figure \ref{fig:OO_compare} shows the distributional comparisons on $d_\mathrm{OO}$. The valid range starts from 2.5 \AA{}, and peaks around 2.75 \AA{} according to $\mathcal{U}_\mathrm{top}$. Hence, $d_\mathrm{OO}$ in range [0, 2.5] \AA{} are considered unphysical. In general, models trained on $\mathcal{V}^{\dagger}$ outperform those trained on $\mathcal{\bar{V}}$.  Similar trends are observed for $d_\mathrm{OH}$ (Fig. \ref{fig:OH_compare}) and $\theta_\mathrm{HOH}$ (Fig. \ref{fig:HOH_compare}).

Figure \ref{fig:Theta_OH_compare} shows the distribution comparisons of the angle between the free OH bond and the surface normal. It's worth noting that for the arising of the peaks around 170$^\circ$, since the water molecules on the bottom layer are missing from the prediction, the OH bond pointing down becomes `free', then contributing to the peak, while this is not the case in calculations from the theoretical structures.  This metric set a high standard, as it requires the model to correctly predict the molecules under the surface. When we put more experimental features into the images in the training data, we generalise a model to adapt to more real features, but it may also lose some sensitivity, making it harder for the underlying molecules to be predicted, since the subtle feature below might be viewed as noise and then be ignored. Models trained on original simulated data tend to capture more of these subtle features, possibly due to their greater sensitivity to weak signals. This trade-off, between generalisation and sensitivity, is also evident in Fig. \ref{fig:v0v1_predictions}, where models trained on $\mathcal{U}$ recover more low-lying hydrogen atoms. Hence, style-translated training improves generalisation but may sacrifice sensitivity, suggesting that future models must carefully balance these two aspects. Confidence scoring for individual atomic predictions would also be beneficial for better interpretation.

Figure \ref{fig:Hbond_compare} shows the joint distributions of the donor–acceptor oxygen distance $d_{\text{O}_\text{d}\text{O}_\text{a}}$ and the hydrogen bond angle $\theta_{\text{O}_\text{d}\text{H}\text{O}_\text{a}}$. It is clear to see that the points obtained from the $F_{\mathcal{V}^{\dagger}}$ models are more concentrated compared to the $F_{\mathcal{\bar{V}}}$ models, indicating better performance in this metric. 

Figure \ref{fig:OrderP_compare} shows the comparisons of the joint distributions of translational ($S_k$) and orientational ($S_g$) tetrahedral order parameters. Models trained on $\mathcal{V}^{\dagger}$ do not significantly outperform those trained on $\mathcal{\bar{V}}$. One reason for that is the generalisation increase with the sacrifice of sensitivity. Another reason is the lower accuracy of distribution estimation, which may be attributed to the limited local environments found in the predicted configurations from the real AFM experiment.

\begin{figure*}[ht!]
    \centering
    \includegraphics[width=\textwidth]{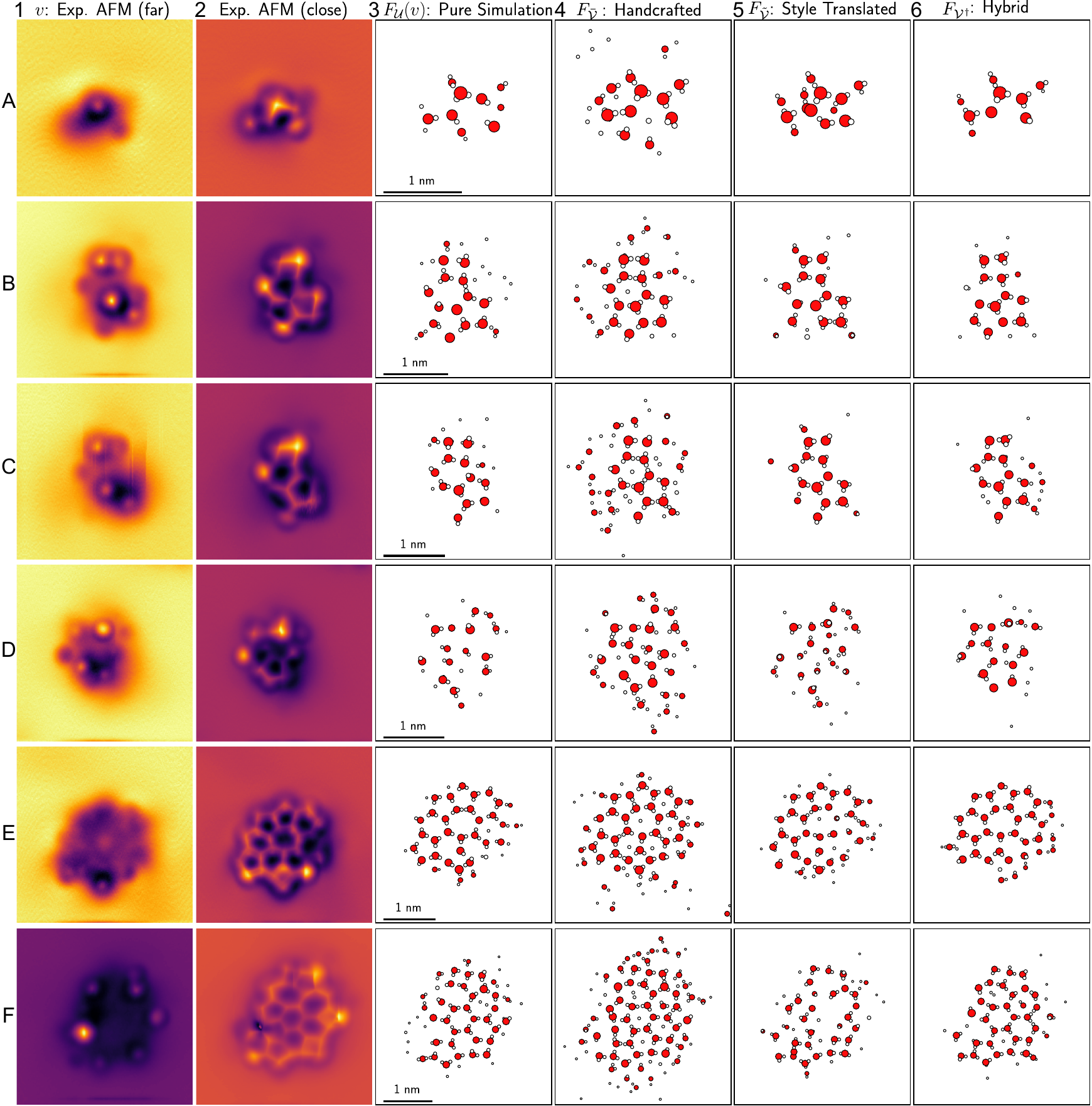}
    \caption{Atomic configuration predictions from 90$^\circ$-rotated experimental AFM images using different structure discovery models.}
    \label{fig:90_predictions}
\end{figure*}

\begin{figure*}[ht!]
    \centering
    \includegraphics[width=\textwidth]{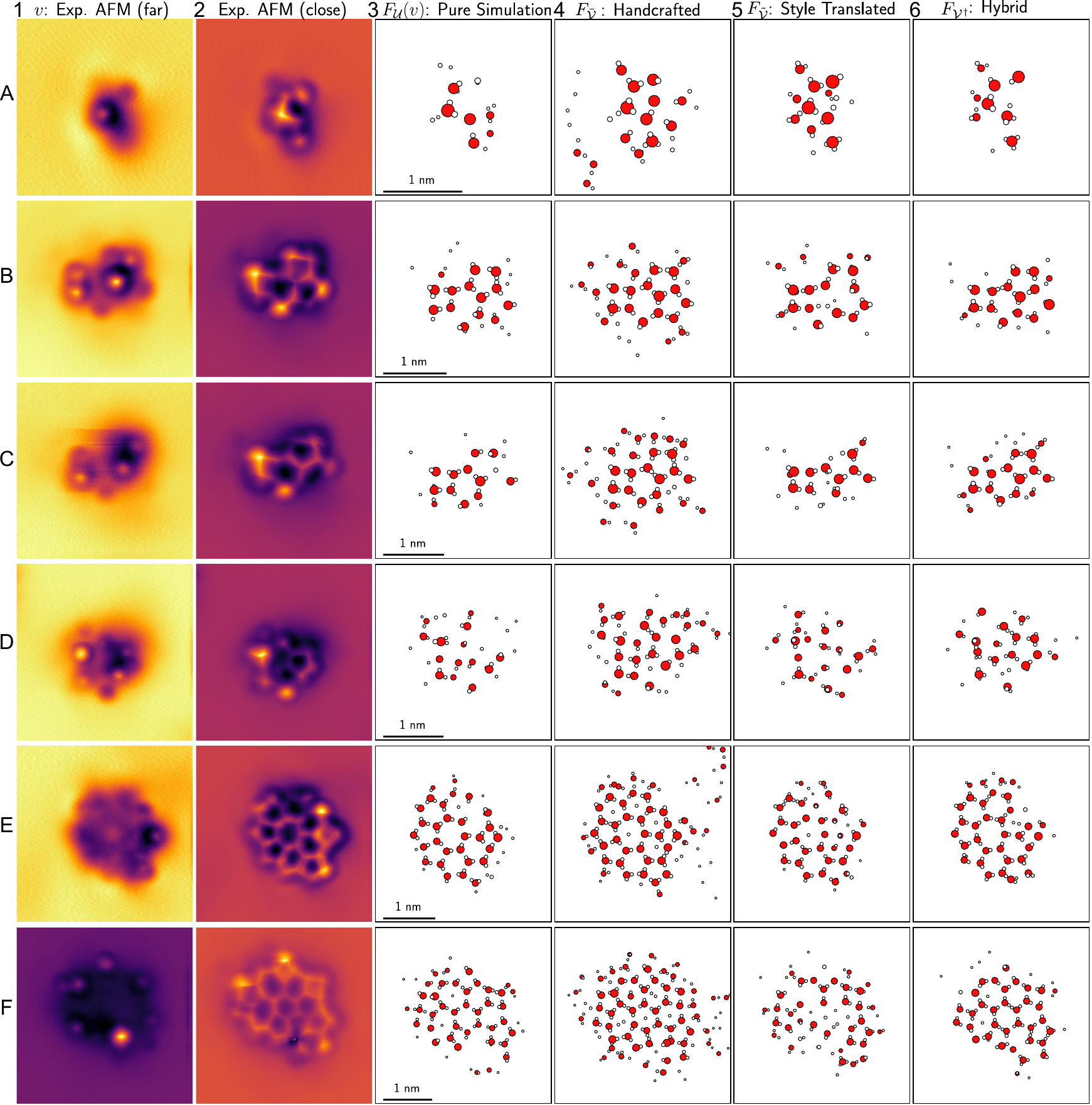}
    \caption{Atomic configuration predictions from 180$^\circ$-rotated experimental AFM images using different structure discovery models.}
    \label{fig:180_predictions}
\end{figure*}

\begin{figure*}[ht!]
    \centering
    \includegraphics[width=\textwidth]{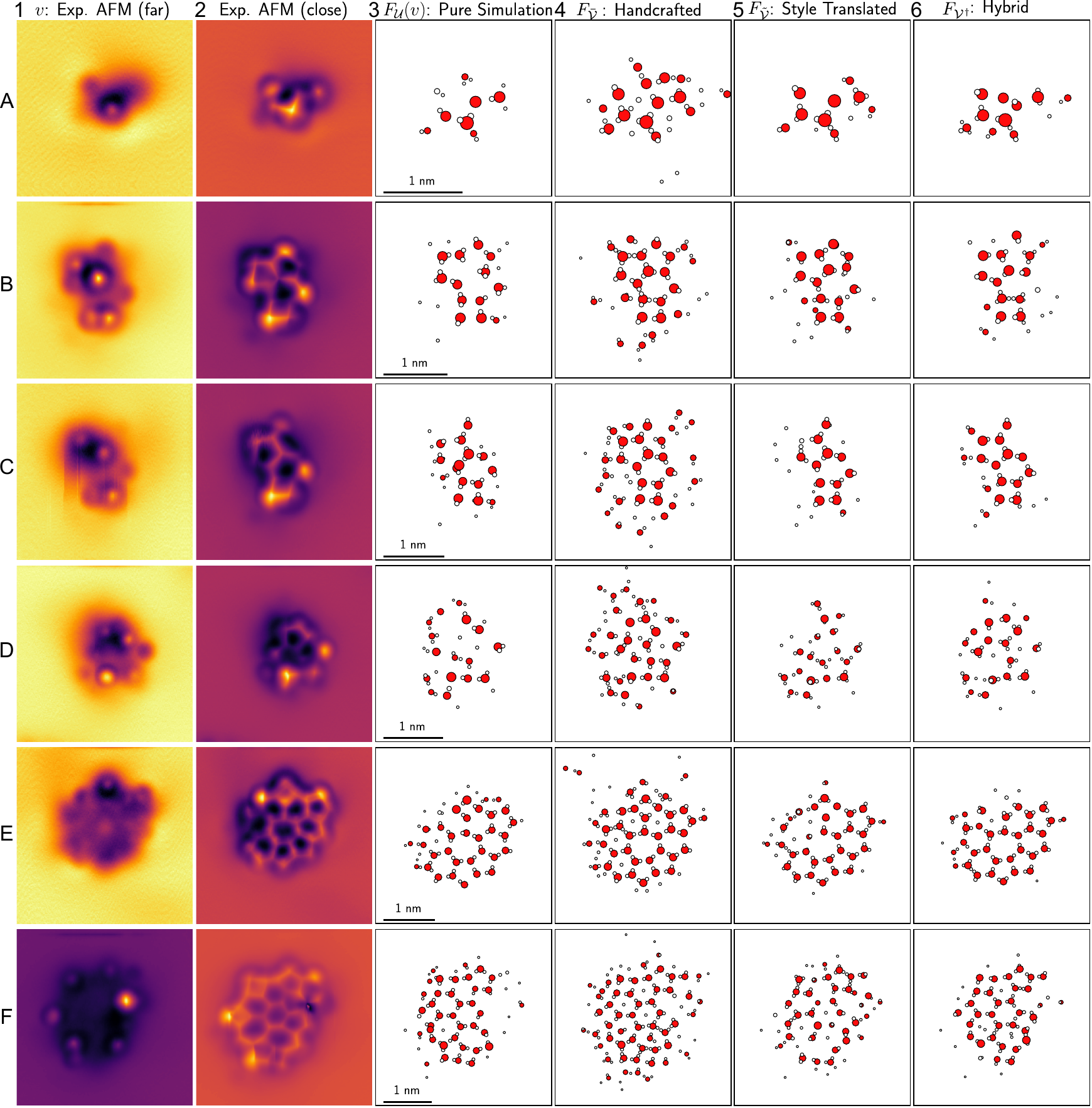}
    \caption{Atomic configuration predictions from 270$^\circ$-rotated experimental AFM images using different structure discovery models.}
    \label{fig:270_predictions}
\end{figure*}

\noindent{\bf Atomic configuration predictions from rotated experimental AFM images.}
Figures \ref{fig:90_predictions}-\ref{fig:270_predictions} show the configuration predictions when rotating the input AFM images.

\end{document}